\documentclass[english,aps,pre,showpacs,floats,twocolumn]{revtex4-1}
\usepackage[T1]{fontenc}
\usepackage[latin9]{inputenc}
\usepackage{amsmath}
\usepackage{amssymb}
\usepackage{graphicx}
\usepackage{esint}

\makeatletter

\newcommand{\lyxdot}{.}

\makeatother

\usepackage{babel}
\begin{document}

\title{Effective anisotropy due to the surface of magnetic nanoparticles}

\author{D. A. Garanin}

\affiliation{Physics Department, Lehman College and Graduate School, The City
University of New York, 250 Bedford Park Boulevard West, Bronx, NY
10468-1589, U.S.A.}

\date{\today}
\begin{abstract}
Analytical solution has been found for the second-order effective
anisotropy of magnetic nanoparticles of a cubic shape due to the surface
anisotropy (SA) of the Néel type. Similarly to the spherical particles,
for the simple cubic lattice the grand-diagonal directions $\left(\pm1,\pm1,\pm1\right)$
are favored by the effective cubic anisotropy but the effect is twice
as strong. Uniaxial core anisotropy and applied magnetic field cause
screening of perturbations from the surface at the distance of the
domain-wall width and reduce the effect of SA near the energy minima.
However, screening disappears near the uniaxial energy barrier, and
the uniform barrier state of larger particles may become unstable.
For these effects the analytical solution is obtained as well, and
the limits of the additive formula with the uniaxial and effective
cubic anisotropies for the particle are established. Thermally-activated
magnetization-switching rates have been computed by the pulse-noise
technique for the stochastic Landau-Lifshitz equation for a system
of spins.
\end{abstract}

\pacs{02.50.Ey, 02.50.-r, 75.78.-n}
\maketitle

\section{Introduction\label{sec:Introduction}}

In magnetic nanoparticles, a significant fraction of atoms belongs
to the surface, and their magnetic properties such as exchange and
anisotropy can be strongly modified. With successes in synthesis of
magnetic particles of a controlled shape, it has become feasible to
investigate the effects of the surface on the magnetic properties
numerically and even analytically. One of the important ingredients
is surface anisotropy (SA) proposed by Néel \cite{nee54} and modeled
microscopically by Victora and MacLaren \cite{vicmac93}. SA arises
due to missing neighbors for the surface spins and breaking the symmetry
of their crystal field and it must be much stronger than a typical
crystallographic anisotropy in the particle's core. However, there
is still not much information on the SA in different materials \cite{chubalhan94,resetal98,chekitokashi99}.
The most common manifestation of the SA is decreasing of the effective
anisotropy of magnetic films with the thickness $d$ as $K_{\mathrm{eff}}=K_{V}+K_{S}/d$,
where $K_{V}$ and $K_{S}$ are volume and surface contributions \cite{chubalhan94,bodmorlin94,resetal98,chekitokashi99}.
It was found that atomic steps on surfaces strongly contribute into
the effective anisotropy \cite{chubalhan94,moronietal03prl}.

In small magnetic clusters, individual spins are tightly bound together
by the exchange interaction, and they form an effective rigid giant
spin with an effective anisotropy dependent on the surface. In particular,
the Néel SA (NSA) was used to model the effective anisotropy of Co
nanoclusters of the form of truncated octahedrons \cite{jametal01,jametal04prb}.
In this case, contributions of different faces and edges partially
cancel each other, leading to a significantly reduced result. For
totally symmetric shapes such as spherical and cubic, the cancellation
of the SA for the rigid cluster's spin is complete.

However, individual spins can deviate from collinearity for stronger
SA and larger particles \cite{dimwys94,dimwys94Cobalt,dimkac02,kacdim02prb}.
Examples of strong noncollinearity are ``throttled'' and ``hedgehog''
spin configurations \cite{dimwys94,dimwys94Cobalt,labetal02,bergeretal08prb}.
Small non-collinearity can be treated perturbatively \cite{garkac03prl,usogre08jap},
that results into the second-order effective anisotropy \cite{garkac03prl}
$K_{\mathrm{eff}}\sim D_{S}^{2}/J$, where $D_{S}$ is the SA and
$J$ is the exchange. For particles with simple cubic (sc) lattice
and spherical shape it was found that the effective second-order anisotropy
has a cubic symmetry with the lowest energy along the grand diagonals
$\left(\pm1,\pm1,\pm1\right)$ of the sc lattice, where the deviations
from collinearity and the resulting energy gain are maximal. The result
scales with the particle's volume as perturbations from the surface
penetrate into the particle's core. Thus in experiments, the second
order effective anisotropy cannot be easily identified with the surface.
If the particle's size becomes too large, deviations from the collinearity
become so strong that the perturbation theory becomes invalid. For
magnetic particles with shapes close to symmetric, the first- and
second-order effective anisotropies can coexist and compete with each
other \cite{yanetal07prb}.

As the SA must be much stronger than the core anisotropy, both first-and
second-order surface terms can compete with the latter. In the simplest
\textit{additive approximation}, for symmetric shapes one can just
add the uniaxial core anisotropy and the cubic second-order surface
anisotropy that leads to complicated energy landscapes \cite{kacbon06prb,yanetal07prb}.
In Refs. \cite{yanetal07prb,yanchu09jpd} it was shown that for the
face-centered (fcc) lattice the sign of the second-order surface anisotropy
is inverted, so that the directions $\left(\pm1,0,0\right)$ etc.,
have the lowest energy. Different temperature dependences of the core
and effective surface anisotropies may cause reorientation transitions
on temperature, as was shown in Ref. \cite{yanchuevacha10jpd} using
the constrained Monte Carlo method \cite{asselinetal10prb}. Additive
effective anisotropy of magnetic particles affects their dynamic properties
such as magnetic resonance \cite{kacsch07epjb} and thermally-activated
switching \cite{dejkackal08jpd,cofdejkal09prb,versabkac14prb}.

As it was mentioned in Ref. \cite{garkac03prl}, in the presence of
the uniaxial core anisotropy $D$, the effect of the surface will
be screened at the distance of the domain-wall width $\delta=a\sqrt{J/(2D)}$
from the surfaces, $a$ being the lattice spacing. Thus the effect
of the surfaces should be reduced for the particle's sizes $L\gtrsim\delta$.
Another effect is the mixing term in the effective anisotropy arising
from both core and surface anisotropies and having another symmetry
\cite{yanetal07prb}.

All analytical and numerical investigations mentioned above were performed
on spherical particles, plus ellipsoidal and truncated octahedron
shapes in Refs. \cite{yanetal07prb,yanchu09jpd}. Analytical solution
for the deviations from collinearity in spherical particles uses the
Green's function for the internal Neumann problem for the Laplace
equation in a sphere. Studying the screening requires solving the
Helmholtz equation for which the Green's function in a sphere is unknown.
For this reason, screening was investigated only perturbatively in
the $L/\delta\ll1$ limit in Ref. \cite{yanetal07prb}. A closed-form
expression for the Green's function of Laplace and Helmholts equations
for cubes and parallelepipeds is unknown. This hampered the investigation
of these shapes, although they are no less important than spherical
and ellipsoidal. For instance, recently Fe nanocubes have been synthesized
\cite{Baibuzetal16ACS-nano}.

Fortunately, for the cubic shape there is an exact analytical solution
that is direct and not using Green's functions. This solution is much
simpler than that for the spherical shape and it allows an extension
for the case of screening by the core anisotropy and by the applied
field. This is the subject of this paper. In particular, it will be
shown that screening is active near the energy minima but becomes
``anti0screening'' closer to the barriers, that leads to stronger
noncollinearities and eventually to the destruction of the quasi-uniform
barrier states.

The plan of the paper is the following. In Sec. \ref{sec:The-model}
the model of classical spins with surface anisotropy is introduced
and the expression for the first-order effective particle's surface
anisotropy is obtained for parallelepipeds. In Sec. \ref{sec:Constrained_minimization}
the method of constrained energy minimization needed to deal with
deviations from spin collinearity in the particle is reviewed and
further developed in comparison to previous publications. This method
is needed for both numerical and analytical work. In Sec. \ref{sec:Numerical-methods}
the numerical implementation of the constrained energy minimization
is discussed. Sec. \ref{sec:Cubic-particle-with} contains the analytical
solution for the second-order effective anisotropy for the cubic particle
with sc lattice in the absence of the core anisotropy and maghetic
field. Sec. \ref{sec:Numerical-results} shows the numerical results
obtained by the constrained energy minimization and their comparizon
with the analytical results. In Sec. \ref{sec:Screening} a more general
analytical solution in the presence of the uniaxial core anisotropy
and the magnetic field is obtained and the effects of screening are
investigated. Sec. \ref{sec:Thermally-activated-escape} presents
the results for the thermally-activated magnetization switching of
the cubic particle considered as a many-spin system.

\section{The model\label{sec:The-model}}

The magnetic particle will be described by the classical spin Hamiltonian
\begin{equation}
\mathcal{H}=-D\sum_{i\in\mathrm{core}}s_{zi}^{2}+\sum_{i\in\mathrm{surface}}\mathcal{H}_{SA,i}-\mathbf{h}\cdot\sum_{i}\mathbf{s}_{i}-\frac{1}{2}\sum_{ij}J_{ij}\mathbf{s}_{i}\cdot\mathbf{s}_{j},\label{Ham}
\end{equation}
where $\mathbf{h}\equiv\mu_{0}\mathbf{H}$ is the magnetic field in
the energy units, $\mathcal{H}_{SA}$ is the surface anisotropy, $J_{ij}$
is the exchange with the coupling $J$ between the neighboring spins
on a sc lattice, $D$ is the core uniaxial-anisotropy constant, and
$|\mathbf{s}_{i}|=1$. The Néel\textquoteright s surface anisotropy
is given by \cite{nee54,vicmac93}
\begin{equation}
{\cal H}_{SA,i}=\frac{1}{2}D_{S}\sum_{j\in nn}\left(\mathbf{s}_{i}\cdot{\bf u}_{ij}\right)^{2},
\end{equation}
where $\mathbf{u}_{ij}$ is the unit vector connecting the surface
site $i$ to its nearest neighbor on site $j$. This anisotropy arises
due to missing nearest neighbors for the surface spins. In particular,
for the simple cubic lattice and $xy$ surfaces (those perpendicular
to $z$ axis), the Néel anisotropy becomes ${\cal H}_{SA,i}=-\frac{1}{2}D_{S}s_{iz}^{2}$
. This means that for $D_{S}>0$ the spins tend to align perpendicularly
to the surface, while for $D_{S}<0$ the surface spins tend to align
parallel to the surface. In a parallelepiped-shaped particle, the
Néel anisotropy on the edges along $z$ axis becomes ${\cal H}_{SA,i}=-\frac{1}{2}D_{S}\left(s_{ix}^{2}+s_{iy}^{2}\right)$
or, equivalently, ${\cal H}_{SA,i}=\frac{1}{2}D_{S}s{}_{iz}^{2}$.
Thus for $D_{S}>0$ the $z$-edge spins tend to align perpendicularly
to $z$ axis. The Néel anisotropy vanishes at the corners and in the
core of the particle.

Spins in particles small enough are tightly bound together by the
exchange and forming an effective giant spin. For a parallelepiped-shaped
particle of the size $N_{z}\times N_{y}\times N_{z}=\mathcal{N}$
the effective Hamiltonian
\begin{equation}
\mathcal{H}_{\mathrm{eff}}=-\mathcal{N}_{\mathrm{core}}Ds_{z}^{2}+\mathcal{H}_{SA}^{(1)}-\mathcal{N}\mathbf{h}\cdot\mathbf{s},\label{Ham_eff_rigid}
\end{equation}
where $\mathcal{N}_{\mathrm{core}}=\left(N_{z}-2\right)\times\left(N_{y}-2\right)\times\left(N_{z}-2\right)$
and $\mathcal{H}_{SA}^{(1)}$ is the sum of contributions from 6 surfaces
and 12 edges
\begin{eqnarray}
\mathcal{H}_{SA}^{(1)} & = & -D_{S}\left[\left(N_{y}N_{z}-4\right)s_{x}^{2}+\left(N_{z}N_{x}-4\right)s_{y}^{2}\right.\nonumber \\
 &  & \qquad\qquad+\left.\left(N_{x}N_{y}-4\right)s_{z}^{2}\right].\label{H_SA_1_result}
\end{eqnarray}
This expression vanishes for $N_{x}=N_{y}=N_{z}=2$ since in this
case there are neither faces nor edges, only corners. For $D_{S}>0,$
the lowest-energy direction is perpendicular to the biggest faces.
For $D_{S}<0,$ the lowest-energy direction is perpendicular to the
smallest faces. For $N_{x}=N_{y}\equiv N_{\bot}$ the model becomes
uniaxial
\begin{equation}
\mathcal{H}_{SA}^{(1)}=-D_{S}N_{\bot}\left(N_{\bot}-N_{z}\right)s_{z}^{2}.\label{H_SA_1_result_uniax}
\end{equation}
The first-order effective anisotropy due to the surface scales with
the surface, thus for large particle sizes $L$ it becomes small as
$1/L$ in comparison to the contribution of the core anisotropy.

\section{Deviations from collinearity and constrained energy minimization\label{sec:Constrained_minimization}}

For the particle of a cubic shape, $\mathcal{H}_{SA}^{(1)}=0$ but
there still is a second-order contribution $\mathcal{H}_{SA}^{(2)}\sim D_{S}^{2}/J$
due to deviation from collinearity generated by the SA. These deviations
depend on the orientation $\boldsymbol{\nu}$ of the particle's magnetization
$\mathbf{m}$, where
\begin{equation}
\mathbf{m}\equiv\frac{1}{\mathcal{N}}\sum_{i}\mathbf{s}_{i},\qquad\boldsymbol{\nu}\equiv\frac{\mathbf{m}}{m}.\label{nu_m_Def}
\end{equation}
Larger deviations correspond to a larger adjustment energy gain, thus
the corresponding directions of $\mathbf{m}$ have lower energy \cite{garkac03prl}.
Deviations from the collinearity are introduced via the formula
\begin{equation}
\mathbf{s}_{i}=\boldsymbol{\nu}\sqrt{1-\boldsymbol{\psi}_{i}^{2}}+\boldsymbol{\psi}_{i}\cong\boldsymbol{\nu}\left(1-\frac{1}{2}\boldsymbol{\psi}_{i}^{2}\right)+\boldsymbol{\psi}_{i},\qquad\sum_{i}\boldsymbol{\psi}_{i}=0.\label{psi_i_Def}
\end{equation}
Below $\boldsymbol{\psi}_{i}$ will be calculated within the linear
approximation.

To define the particle's energy for different $\mathbf{m}$ directions
$\boldsymbol{\nu}$, one has to constrain the latter. This can be
done by using the method of Lagrange multipliers \cite{garkac03prl,kacbon06prb,yanetal07prb,pazgarchu08pa}
in which one minimizes the function

\begin{equation}
\mathcal{F}\equiv\mathcal{H}-\mathcal{N}\boldsymbol{\lambda}\cdot\left(\boldsymbol{\nu}-\boldsymbol{\nu}_{0}\right),\label{Lagrange}
\end{equation}
where $\boldsymbol{\nu}_{0}$ is the preset direction, $|\boldsymbol{\nu}_{0}|=1$.
The constrained equilibrium solution satisfies the equations
\begin{equation}
\mathbf{s}_{i}\times\frac{\partial\mathcal{F}}{\partial\mathbf{s}_{i}}=0,\qquad\frac{\partial\mathcal{F}}{\partial\boldsymbol{\lambda}}=-\mathcal{N}\left(\boldsymbol{\nu}-\boldsymbol{\nu}_{0}\right)=0.\label{constrained_equations}
\end{equation}
From the second equation follows $\boldsymbol{\nu}=\boldsymbol{\nu}_{0}$.
In the first equation
\begin{equation}
-\frac{\partial\mathcal{F}}{\partial\mathbf{s}_{i}}=\mathbf{h}_{\mathrm{eff},i}+\mathbf{h}_{\lambda},\qquad\mathbf{h}_{\mathrm{eff},i}\equiv-\frac{\partial\mathcal{H}}{\partial\mathbf{s}_{i}}\label{h_eff_Def}
\end{equation}
and the constraint field is uniform and given by
\begin{equation}
\mathbf{h}_{\lambda}\equiv\mathcal{N}\frac{\partial\left(\boldsymbol{\lambda}\cdot\boldsymbol{\nu}\right)}{\partial\mathbf{s}_{i}}=\frac{1}{m}\left[\boldsymbol{\lambda}-\boldsymbol{\nu}\cdot\left(\boldsymbol{\nu}\cdot\boldsymbol{\lambda}\right)\right].\label{h_constr_via_lam}
\end{equation}
 Note that $\mathbf{h}_{\lambda}$ is perpendicular to $\mathbf{m}$
since it constrains only its direction $\boldsymbol{\nu}$, leaving
its magnitude $m$ free to change.

Analytically, the constraint field can be found at zero order in $\boldsymbol{\psi}_{i}$,
considering the rigid particle's spin and averaging the effective
field over the particle to get the contribution of the surface. Thus
the first of equations (\ref{constrained_equations}) becomes
\begin{equation}
\boldsymbol{\nu}\times\left(\mathbf{\bar{h}}_{\mathrm{eff}}+\mathbf{h}_{\lambda}\right)=0,\label{eq_equil_with_h_constr}
\end{equation}
where from Eq. (\ref{Ham_eff_rigid}) one obtains
\begin{equation}
\mathbf{\bar{h}}_{\mathrm{eff}}=\mathbf{h}+2\tilde{D}\nu_{z}\mathbf{e}_{z}+\mathbf{h}_{SA},\qquad\tilde{D}\equiv\frac{\mathcal{N}_{\mathrm{core}}}{\mathcal{N}}D\label{h_eff_average}
\end{equation}
and
\begin{equation}
\mathbf{h}_{SA}\equiv-\frac{1}{\mathcal{N}}\frac{\partial\mathcal{H}_{SA}^{(1)}\left(\boldsymbol{\nu}\right)}{\partial\boldsymbol{\nu}}.\label{h_SA}
\end{equation}
In particular, for $N_{x}=N_{y}\equiv N_{\bot}$ one obtains
\begin{equation}
\mathbf{h}_{SA}=2D_{S}\frac{N_{\bot}-N_{z}}{N_{\bot}N_{z}}s_{s}\mathbf{e}_{z}.\label{h_SA_uniaxial}
\end{equation}
Since $\mathbf{h}_{\lambda}$ is perpendicular to $\boldsymbol{\nu}$,
the solution of Eq. (\ref{eq_equil_with_h_constr}) is
\begin{equation}
\mathbf{h}_{\lambda}=-\mathbf{\bar{h}}_{\mathrm{eff}}+\boldsymbol{\nu}\left(\boldsymbol{\nu}\cdot\mathbf{\bar{h}}_{\mathrm{eff}}\right).\label{h_constr_via_h_eff}
\end{equation}

\section{Numerical methods\label{sec:Numerical-methods}}

\begin{figure}
\begin{centering}
\includegraphics[width=9cm]{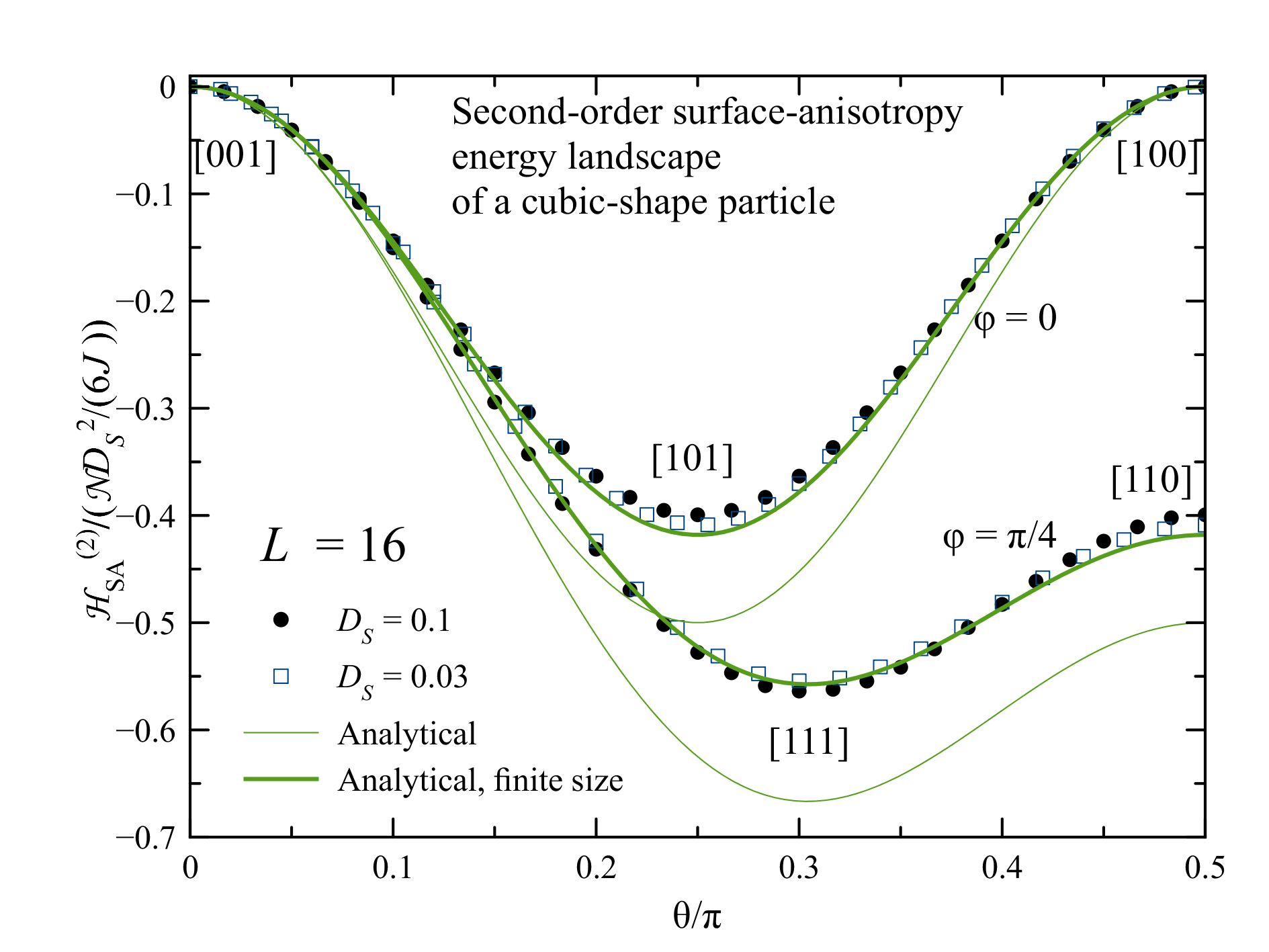}
\par\end{centering}
\begin{centering}
\includegraphics[width=9cm]{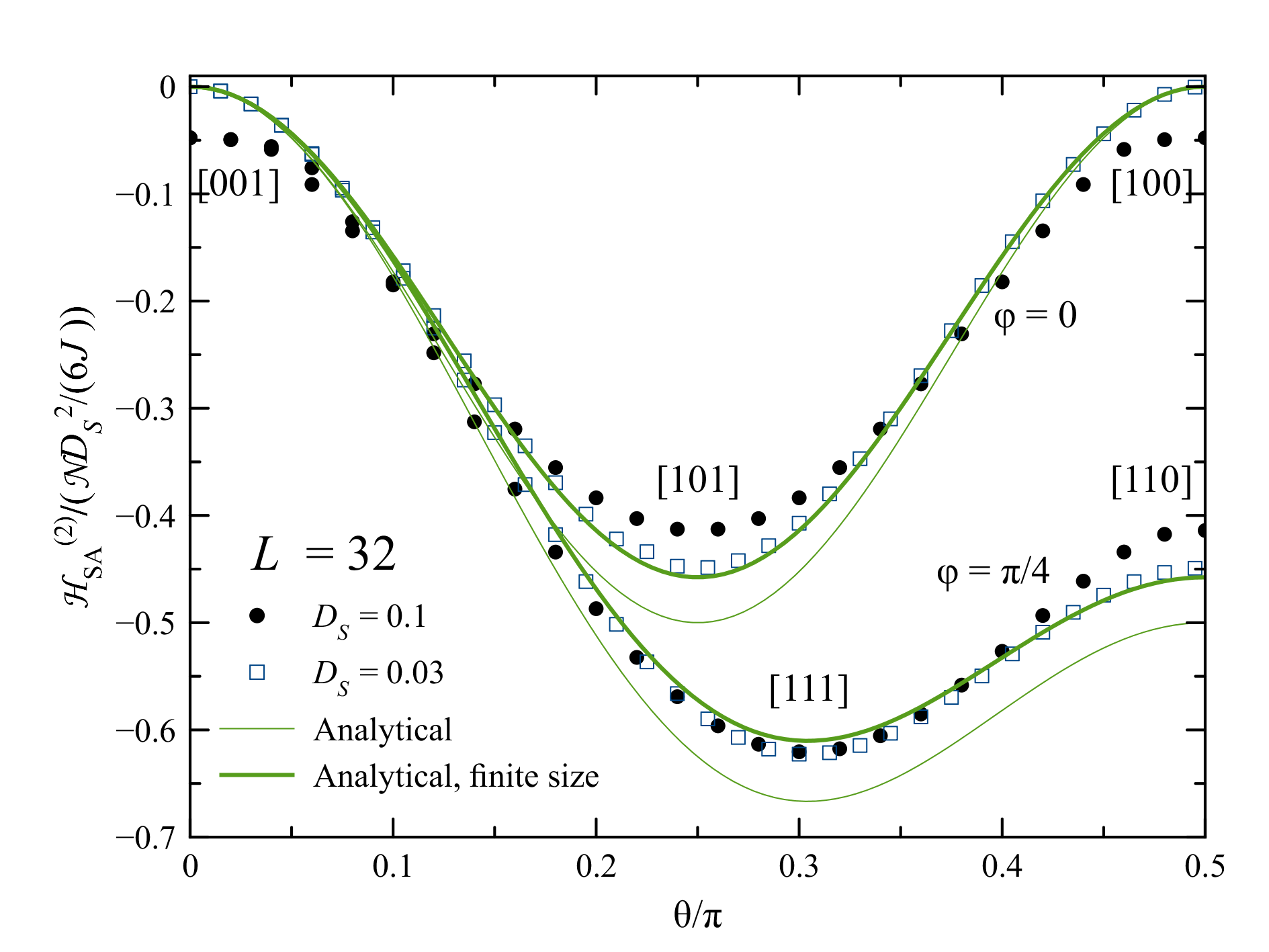}
\par\end{centering}
\caption{Energy landscape for the cubic particle with the surface anisotropy:
Numerical and analytical results for the particle sizes $L=16$, 32
and the SA strenghts $D_{S}=0.1,$ 0.03. ``Analytical'' curves use
Eq. (\ref{H_SA^(2)_DS_result}) while ``Analytical, finite size''
curves use Eq. (\ref{H_SA^(2)_DS_result}) with the additional factor
$(1-0.7/L)^{4}$.}

\label{Fig-Ered_vs_theta_DV=00003D0}
\end{figure}
\begin{figure}
\begin{centering}
\includegraphics[width=9cm]{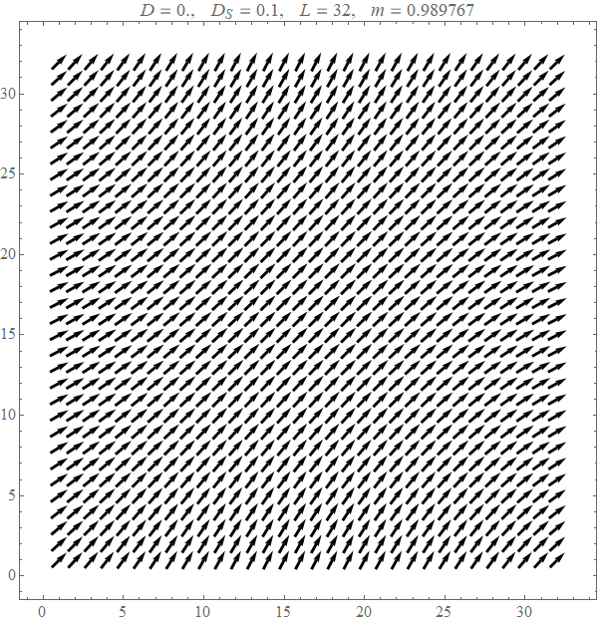}
\par\end{centering}
\begin{centering}
\includegraphics[width=9cm]{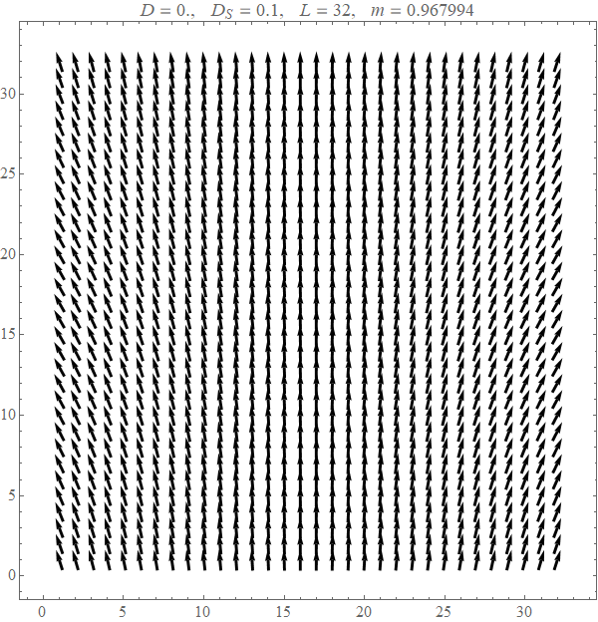}
\par\end{centering}
\caption{Spin structures in a cubic particle of the size $L=32$ for $D_{S}=0.1$,
scans through middle of the particle. Upper panel: particle's magnetization
direction $\left(1,1,0\right)$ etc. Spins slightly canted to lower
the system's energy in accordance with the analytical solution, Eq.
(\ref{psi_solution_DS_only_cube}). Lower panel: particle's magnetization
direction $\left(0,0,1\right)$. Spins slightly turned toward the
directions perpendicular to the right and left surfaces as the result
of the instability of the collinear state (for $L=16$ spins are still
strictly collinear).}

\label{Fig_spins_L=00003D32_DS=00003D0.1}
\end{figure}

One practical method of numerically solving Eq. (\ref{constrained_equations})
is the method of relaxation in which the evolution equations
\begin{eqnarray}
\mathbf{\dot{s}}_{i} & = & -\alpha\mathbf{s}_{i}\times\left[\mathbf{s}_{i}\times\left(\mathbf{h}_{\mathrm{eff},i}+\mathbf{h}_{\lambda}\right)\right]\nonumber \\
\dot{\boldsymbol{\lambda}} & = & \frac{\alpha_{\lambda}}{\mathcal{N}}\frac{\partial\mathcal{F}}{\partial\boldsymbol{\lambda}}=-\alpha_{\lambda}\left(\boldsymbol{\nu}-\boldsymbol{\nu}_{0}\right)
\end{eqnarray}
with relaxation constants $\alpha$ and $\alpha_{\lambda}$ are solved
\cite{garkac03prl,kacbon06prb,yanetal07prb}. A faster method is a
combination of the field alignment and overrelaxation used in Refs.
\cite{garchupro13epl,garchupro13prb,progarchu14prl} for finding local
energy minima in magnetic systems with quenched randomness. In this
method, all spins $\mathbf{s}_{i}$ are updated consequtively by the
field alignment $\mathbf{s}_{i,\mathrm{new}}=\mathbf{h}_{\mathrm{eff},i}/\left|\mathbf{h}_{\mathrm{eff},i}\right|$
or the overrelaxation $\mathbf{s}_{i,\mathrm{new}}=2\left(\mathbf{s}_{i,\mathrm{old}}\cdot\mathbf{h}_{\mathrm{eff},i}\right)\mathbf{h}_{\mathrm{eff},i}/h_{\mathrm{eff},i}^{2}-\mathbf{s}_{i,\mathrm{old}}$
with the probabilities $\alpha$ and $1-\alpha$, respectively. The
first procedure is pseudorelaxation while the second is pseudodynamics
flipping the spins by 180$^{\circ}$ around the effective field. The
highest efficiency of this method is achieved in the underdamped regime
$\alpha=0.1\div0.01$. For the constrained minimization here, one
has to replace $\mathbf{h}_{\mathrm{eff},i}\Rightarrow\mathbf{h}_{\mathrm{eff},i}+\mathbf{h}_{\lambda}$
and add the iteration $\boldsymbol{\lambda}_{\mathrm{new}}=\boldsymbol{\lambda}_{\mathrm{old}}-\alpha_{\lambda}\left(\boldsymbol{\nu}-\boldsymbol{\nu}_{0}\right)$
at the end of each full-system spin update. The spin updates within
this method are parallelizable that leads to a significant speed-up.

The method of costrained energy minimization works well if the spin
noncollinearity is small enough. In this case the particle's energy
is a nice one-valued function showing minima for the grand-diagonal
directions for spherical \cite{garkac03prl,kacbon06prb,yanetal07prb}
and cubic particles with a sc lattice. For larger $D_{S}$ and $L$,
the solution looks distorted and can become multi-valued. Further
increase of these parameters may results in the loss of convergence.
The physical reason for this is that the particle is no more in the
single-domain state that is a prerequisite for the method's validity.
In particular, even in the absence of the SA, large particles are
overcoming the energy barrier due to the uniaxial anisotropy via a
non-uniform rotation in which a domain wall is moving across the particle.
Onset of this regime leads to the failure of the constrained minimization
method.

In the numerical work, $J\Rightarrow1$ and $a\Rightarrow1$ are set.

\section{Cubic particle with surface anisotropy only\label{sec:Cubic-particle-with}}

\begin{figure}
\begin{centering}
\includegraphics[width=9cm]{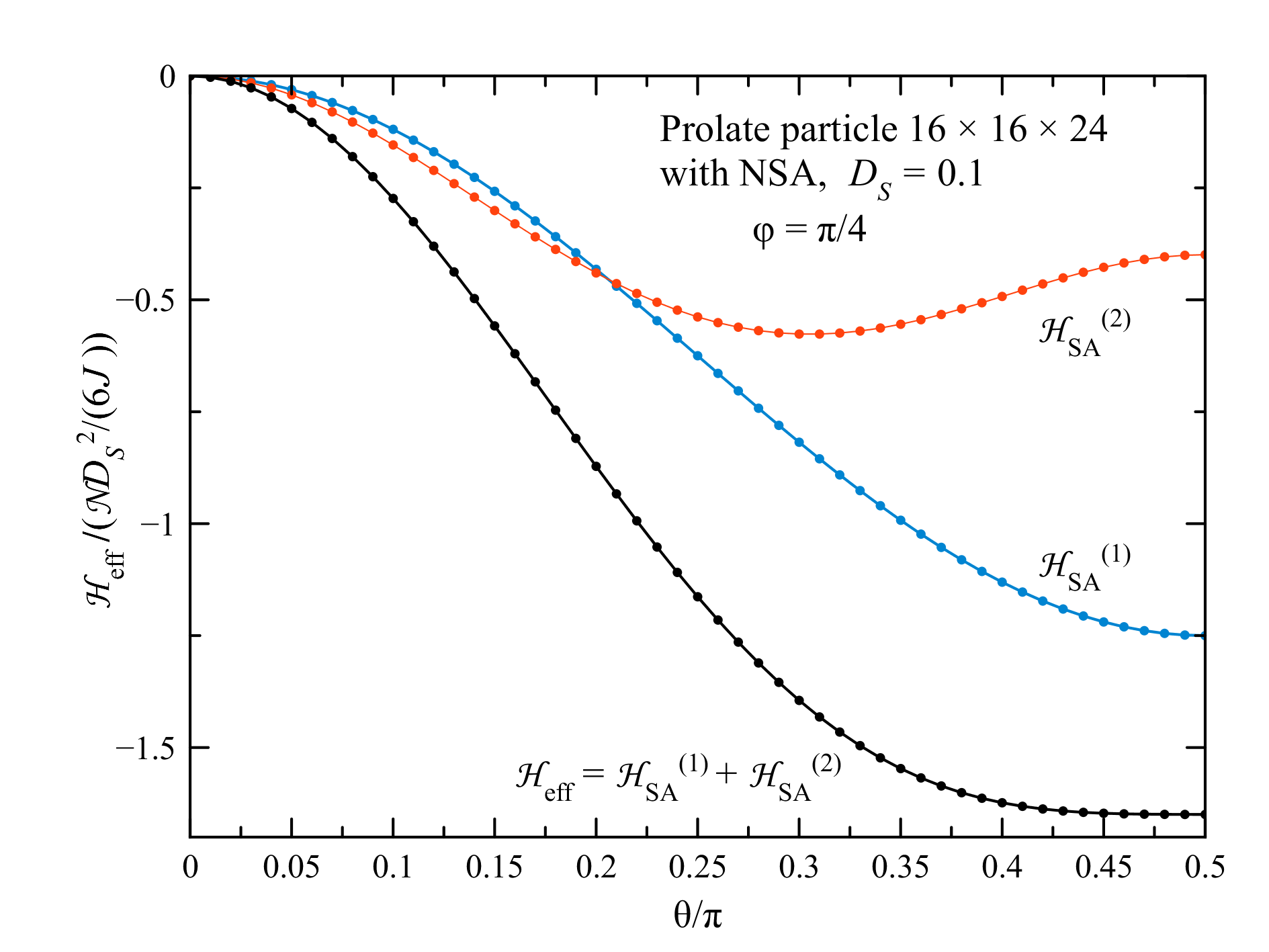}
\par\end{centering}
\begin{centering}
\includegraphics[width=9cm]{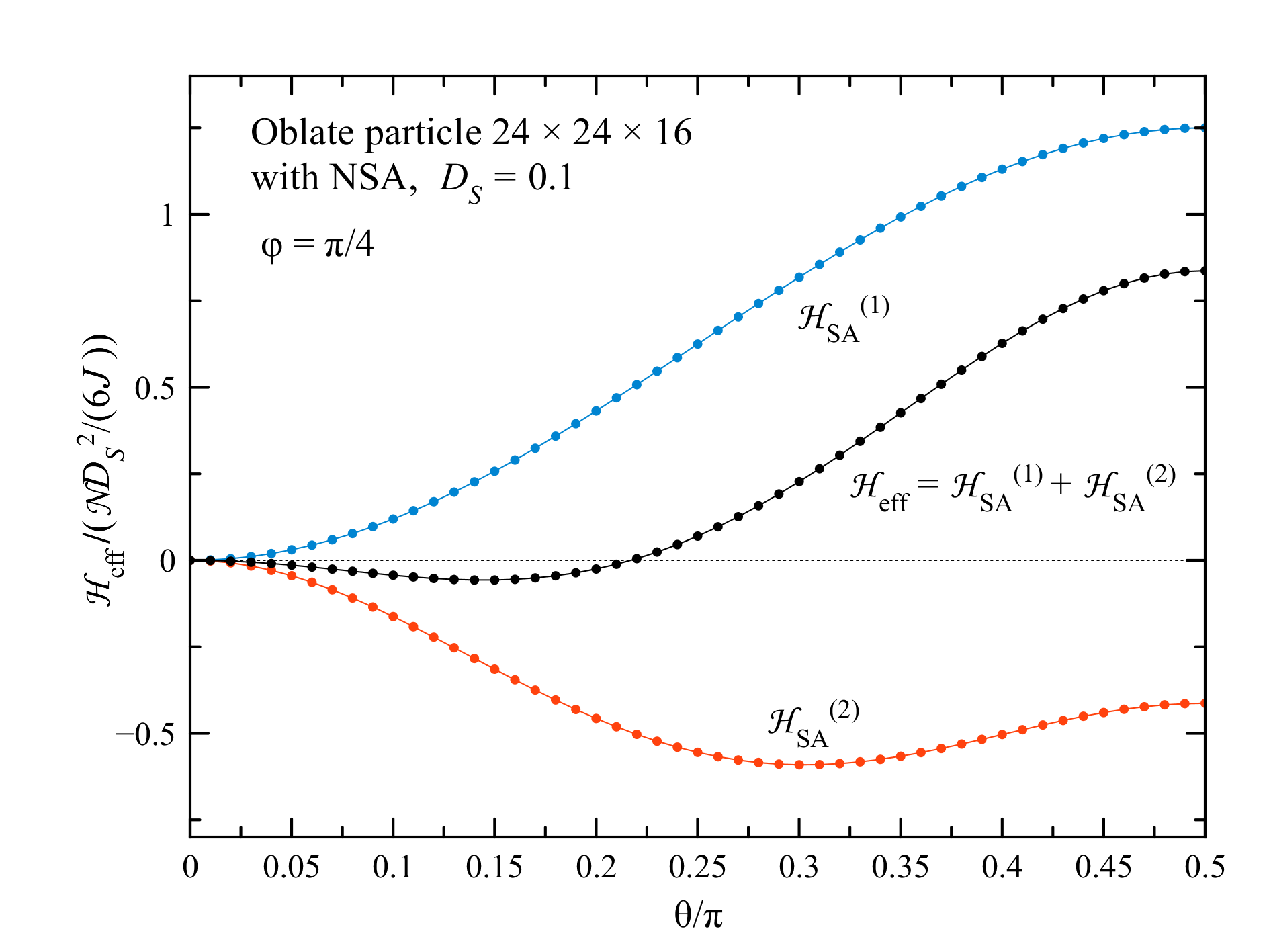}
\par\end{centering}
\caption{Energy landscape for prolate and oblate particles with SA. The second-order
effective anisotropy computed as $\mathcal{H}_{SA}^{(2)}=\mathcal{H}_{\mathrm{eff}}-\mathcal{H}_{SA}^{(1)}$
is comparable with the first-order one. Upper panel: prolate particle.
Lower panel: oblate particle.}

\label{Fig-Ered_vs_theta_oblate_and_prolate}
\end{figure}
\begin{figure}
\begin{centering}
\includegraphics[width=9cm]{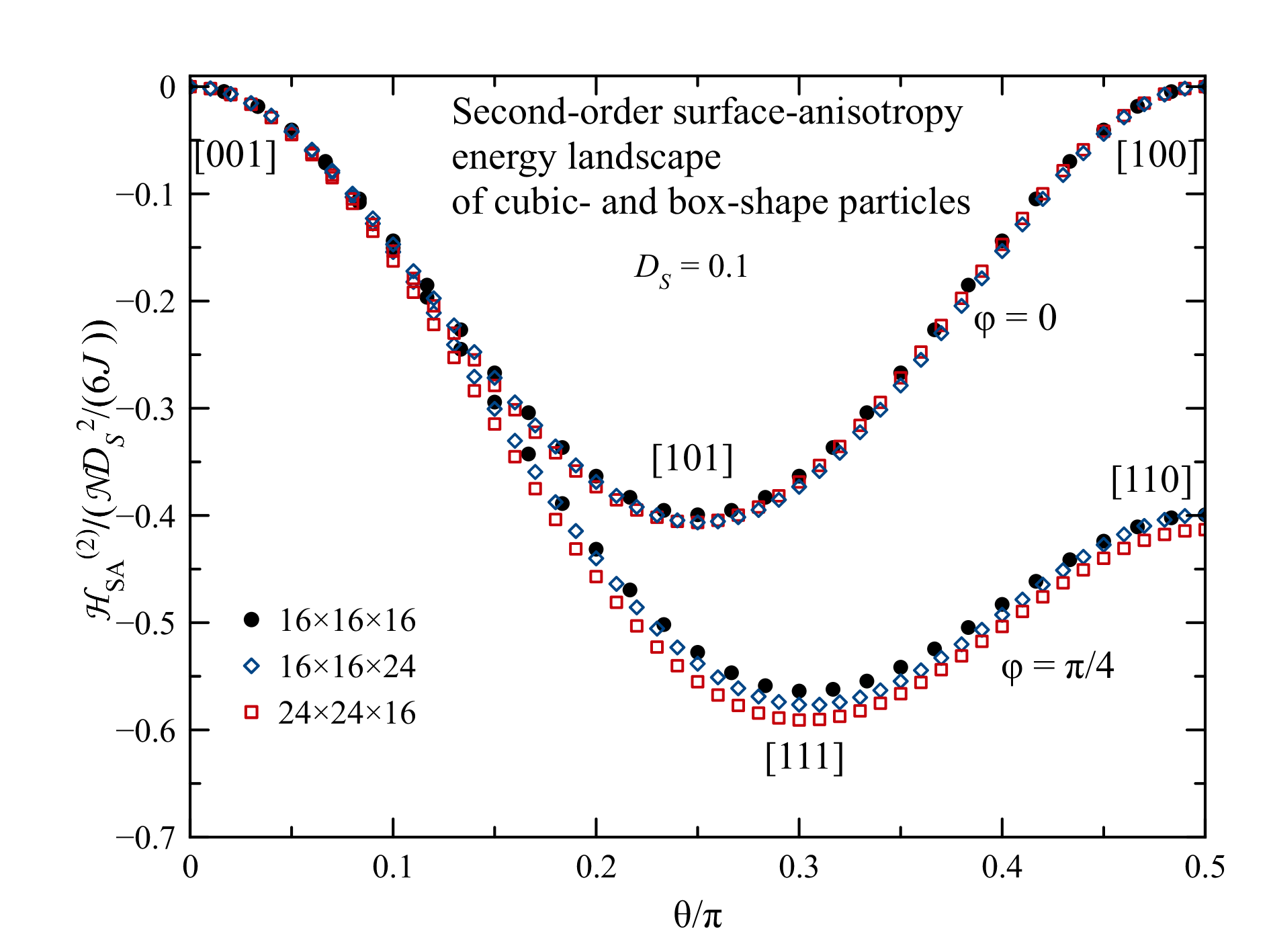}
\par\end{centering}
\caption{Second-order effective anisotropy for cubic, prolate, and oblate particles
with SA. The result is practically independent of the particle's aspect
ratio.}

\label{Fig-Ered_vs_theta_DS=00003D0.1_cubic_and_box}
\end{figure}

The analytical solution for the second-order effective surface anisotropy
is simpler for the cubic-shaped particle with the SA only. In this
case, in the rigid-spin approximation, there is no effective field
acting on the spin, so according to Eq. (\ref{h_constr_via_h_eff})
there is no constrait field $\mathbf{h}_{\lambda}$ as well. (A similar
problem was considered in Ref. \cite{usogre08jap}, however, no effective
cubic anisotropy was obtained.) In the continuous approximation, the
particle's energy has the form
\begin{equation}
\mathcal{H}=\frac{1}{a^{3}}\intop dV\left[\frac{1}{2}a^{2}J\left(\frac{\partial s_{\alpha}}{\partial\mathbf{r}}\right)^{2}-\frac{1}{2}D_{S}a\delta_{S}\left(\mathbf{n}\cdot\mathbf{s}\right)^{2}\right]\label{Ham_continuous_SA}
\end{equation}
with summation over the repeated $\alpha$. Here $\delta_{S}$ is
the surface $\delta$-function and $\mathbf{n}$ is the outer normal
to the surface. Minimizing this energy leads to the equation
\begin{equation}
\mathbf{s}\times\Delta\mathbf{s}=0\label{equilibrium_eq}
\end{equation}
with the boundary condition
\begin{equation}
\mathbf{s}\times\left(a\frac{\partial\mathbf{s}}{\partial r_{\alpha}}n_{\alpha}-\frac{D_{S}}{J}\left(\mathbf{n}\cdot\mathbf{s}\right)\mathbf{n}\right)=0\label{Boundary_condition_general}
\end{equation}
at the surfaces.

Now, considering $D_{S}/J$ as small and starting from a collinear
state of a fixed direction $\boldsymbol{\nu}$, one can consider surface-induced
deviations $\boldsymbol{\boldsymbol{\psi}}$ from this state,
\begin{equation}
\mathbf{s}(\mathbf{r})=\boldsymbol{\nu}\sqrt{1-\boldsymbol{\boldsymbol{\psi}}^{2}(\mathbf{r})}+\boldsymbol{\boldsymbol{\psi}}(\mathbf{r})\cong\boldsymbol{\nu}\left[1-\frac{1}{2}\boldsymbol{\boldsymbol{\psi}}^{2}(\mathbf{r})\right]+\boldsymbol{\boldsymbol{\psi}}(\mathbf{r}),\label{psi_Def}
\end{equation}
where $\mathbf{s}(\mathbf{r})\cdot\boldsymbol{\boldsymbol{\psi}}(\mathbf{r})=0$
and $\boldsymbol{\boldsymbol{\psi}}$ satisfies the sum rule
\begin{equation}
\int d^{3}r\boldsymbol{\boldsymbol{\psi}}(\mathbf{r})=0.\label{psi_sum_rule}
\end{equation}
Below $\boldsymbol{\psi}$ will be found within the linear approximation,
whereas the quadratic term will be used to calculate the decrease
of the particle's magnetization $m$ \textendash{} the magnetization
deficit. The equation for $\boldsymbol{\boldsymbol{\psi}}$ becomes
\begin{equation}
\Delta\boldsymbol{\boldsymbol{\psi}}=0,\qquad a\frac{\partial\boldsymbol{\boldsymbol{\psi}}}{\partial r_{\alpha}}n_{\alpha}=\frac{D_{S}}{J}\mathbf{f}\left(\boldsymbol{\nu},\mathbf{n}\right),\label{psi_eq}
\end{equation}
where
\begin{equation}
\mathbf{f}\left(\boldsymbol{\nu},\mathbf{n}\right)\equiv\left(\mathbf{n}\cdot\boldsymbol{\nu}\right)\left[\mathbf{n}-\left(\mathbf{n}\cdot\mathbf{\boldsymbol{\nu}}\right)\mathbf{\boldsymbol{\nu}}\right]
\end{equation}
is perpendicular to $\boldsymbol{\nu}$ and vanishes if $\mathbf{\boldsymbol{\nu}}$
is perpendicular to any particle's face, $\mathbf{n}\cdot\mathbf{\boldsymbol{\nu}}=\pm1$.
For the parallelepiped of linear sizes $L_{x,y,z}$, the boundary
conditions become
\begin{equation}
\pm a\left.\frac{\partial\boldsymbol{\boldsymbol{\psi}}}{\partial x}\right|_{x=\pm L_{x}/2}=\frac{D_{S}}{J}\mathbf{f}\left(\boldsymbol{\nu},\mathbf{n}\right)\label{psi_boundary_conditions}
\end{equation}
etc. At the opposite faces of the particle $\mathbf{f}\left(\boldsymbol{\nu},\mathbf{n}\right)$
is the same as it is quadratic in $\mathbf{n}$. At different faces,
$\mathbf{f}\left(\boldsymbol{\nu},\mathbf{n}\right)$ are, in general,
different. The explicit values are given by
\begin{equation}
\mathbf{f}\left(\boldsymbol{\nu},\mathbf{n}\right)=\begin{cases}
\nu_{x}\left(\mathbf{e}_{x}-\nu_{x}\boldsymbol{\nu}\right), & x=\pm L_{x}/2\\
\nu_{y}\left(\mathbf{e}_{y}-\nu_{y}\boldsymbol{\nu}\right), & y=\pm L_{y}/2\\
\nu_{z}\left(\mathbf{e}_{z}-\nu_{z}\boldsymbol{\nu}\right), & z=\pm L_{z}/2.
\end{cases}\label{f_on_faces}
\end{equation}
For the cube of linear size $L$, one can search for the solution
in the form
\begin{equation}
\boldsymbol{\psi}(\mathbf{r})=\frac{D_{S}}{LaJ}\left(\mathbf{C}_{x}x^{2}+\mathbf{C}_{y}y^{2}+\mathbf{C}_{z}z^{2}\right)\label{psi_solution_DS_only_cube}
\end{equation}
that satisfies the Laplace equation and at the same time Eq. (\ref{psi_sum_rule})
for
\begin{equation}
\mathbf{C}_{x}+\mathbf{C}_{y}+\mathbf{C}_{z}=0.\label{C_condition_Laplace}
\end{equation}
From the boundary conditions above one finds
\begin{equation}
\mathbf{C}_{x,y,z}=\left.\mathbf{f}\left(\boldsymbol{\nu},\mathbf{n}\right)\right|_{x,y,z=\pm L/2},
\end{equation}
the values from Eq. (\ref{f_on_faces}). One can check that this solution
satisfies Eq. (\ref{C_condition_Laplace}).

The maximal value of $\psi$ reached at the surfaces of the particle
should be small,
\begin{equation}
\psi\sim\frac{L}{a}\frac{D_{S}}{J}\ll1,\label{psi_applicability}
\end{equation}
that defines the applicability range of the linearization. For $\nu_{x}=\nu_{y}=\nu_{z}=1/\sqrt{3}$
at the center of the face $x=L/2$, $y=z=0$ one has $6\sqrt{2}\simeq8.5$
in the denominator of this formula, thus the applicability condition
is milder than above. On the other hand, there are instabilities of
the found states at larger $L$ and $D_{S}$ that were observed numerically
but haven't been yet worked out analytically. These instabilities
also limit the applicability of the method.

Now the particle's magnetization can be computed using the quadratic
terms in Eq. (\ref{psi_Def}) as
\begin{equation}
\mathbf{m}=\frac{1}{V}\iiintop_{0}^{L}dxdydz\left\{ \boldsymbol{\nu}\left[1-\frac{1}{2}\boldsymbol{\boldsymbol{\psi}}^{2}(\mathbf{r})\right]+\boldsymbol{\boldsymbol{\psi}}(\mathbf{r})\right\} .
\end{equation}
Here the linear term vanishes while the quadratic term yields
\begin{equation}
\mathbf{m}=\boldsymbol{\nu}\left[1-\frac{1}{360}\left(\frac{L}{a}\frac{D_{S}}{J}\right)^{2}\left(1-\nu_{x}^{4}-\nu_{y}^{4}-\nu_{z}^{4}\right)\right].\label{m_DS_result}
\end{equation}
Clearly, the magnetization deficit vanishes if $\mathbf{\boldsymbol{\nu}}$
is perpendicular to any particle's face and reaches its maximum for
the grand-diagonal directions $\left(\pm1,\pm1,\pm1\right)$. Again,
the small coefficient in this formula suggests that the applicability
condition for the linearization method is milder than given by Eq.
(\ref{psi_applicability}).

The energy of the particle for the state found above at the lowest,
quadratic, order in $D_{S}$ becomes
\begin{equation}
\mathcal{H}_{SA}^{(2)}=\intop\frac{dV}{a^{3}}\left[\frac{1}{2}a^{2}J\left(\frac{\partial\psi_{\alpha}}{\partial\mathbf{r}}\right)^{2}-D_{S}a\delta_{S}\left(\mathbf{n}\cdot\boldsymbol{\nu}\right)\left(\mathbf{n}\cdot\boldsymbol{\psi}\right)\right],\label{E_psi_continuous}
\end{equation}
i.e., $\mathcal{H}_{SA}^{(2)}=E_{ex}+E_{DS}$. After integration one
obtains
\begin{equation}
\mathcal{H}_{SA}^{(2)}=-\frac{\mathcal{N}D_{S}^{2}}{6J}\left(1-\nu_{x}^{4}-\nu_{y}^{4}-v_{z}^{4}\right),\label{H_SA^(2)_DS_result}
\end{equation}
whereas $E_{ex}=-\mathcal{H}_{SA}^{(2)}>0$ and $E_{DS}=2\mathcal{H}_{SA}^{(2)}<0$.
This result is similar to that for the spherical particle \cite{garkac03prl}
and differs from it by the missing factor $\kappa\simeq0.53.$ Adding
the first-order effective particle's Hamiltonian, Eq. (\ref{Ham_eff_rigid}),
one obtains
\begin{equation}
\mathcal{H}_{\mathrm{eff}}=-\mathcal{N}_{\mathrm{core}}Ds_{z}^{2}-\mathcal{N}\mathbf{h}\cdot\mathbf{s}+\mathcal{H}_{SA}^{(1)}+\mathcal{H}_{SA}^{(2)},\label{Ham_eff_1+2}
\end{equation}
where $\mathcal{H}_{SA}^{(1,2)}$ are given by Eqs. (\ref{H_SA_1_result})
and (\ref{H_SA^(2)_DS_result}), respectively. This \textit{additive}
approximation does not take into account screening and is good for
not too large particle's sizes $L$.

\section{Numerical results\label{sec:Numerical-results}}

\begin{figure}
\begin{centering}
\includegraphics[width=9cm]{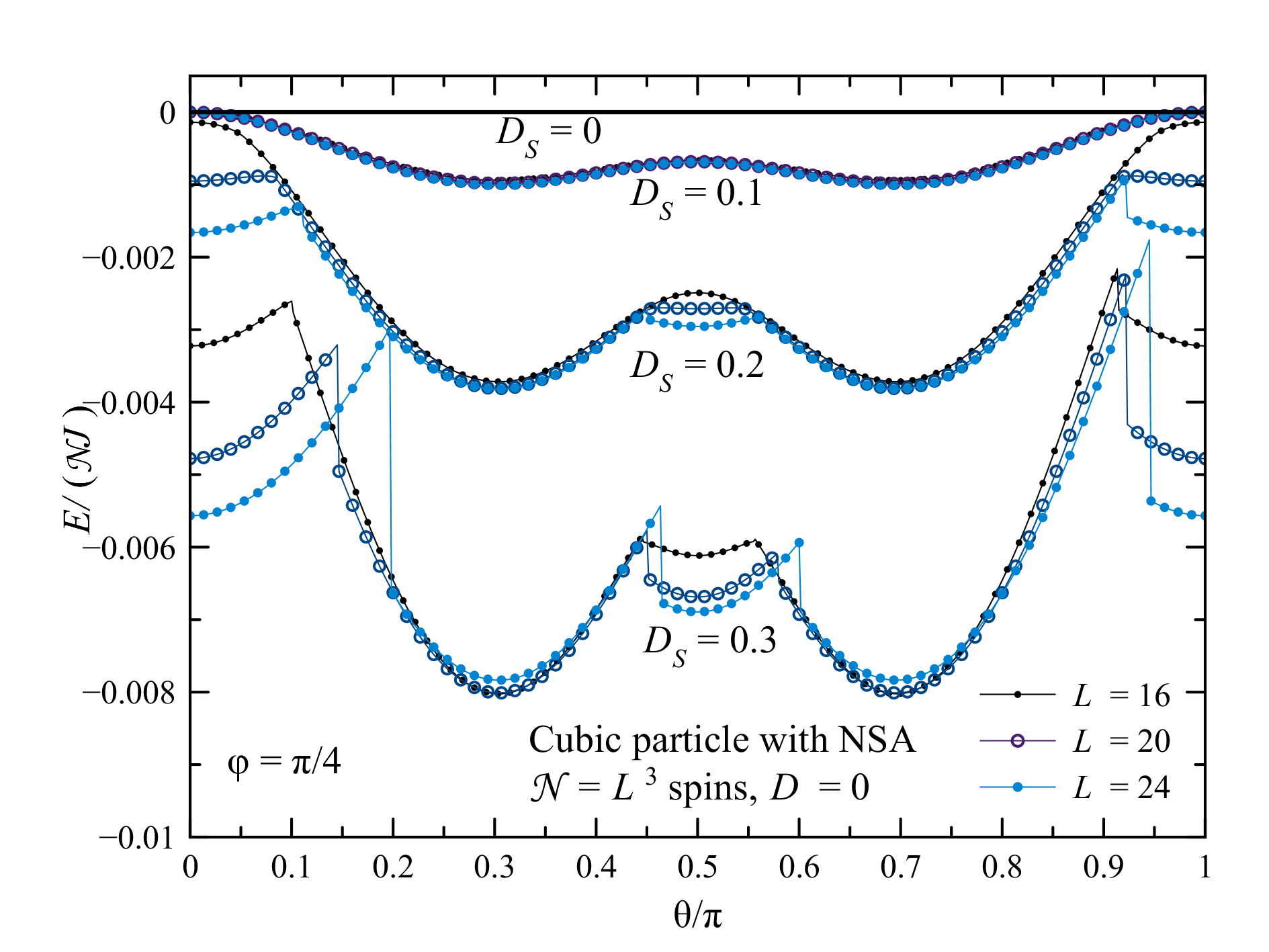}
\par\end{centering}
\begin{centering}
\includegraphics[width=9cm]{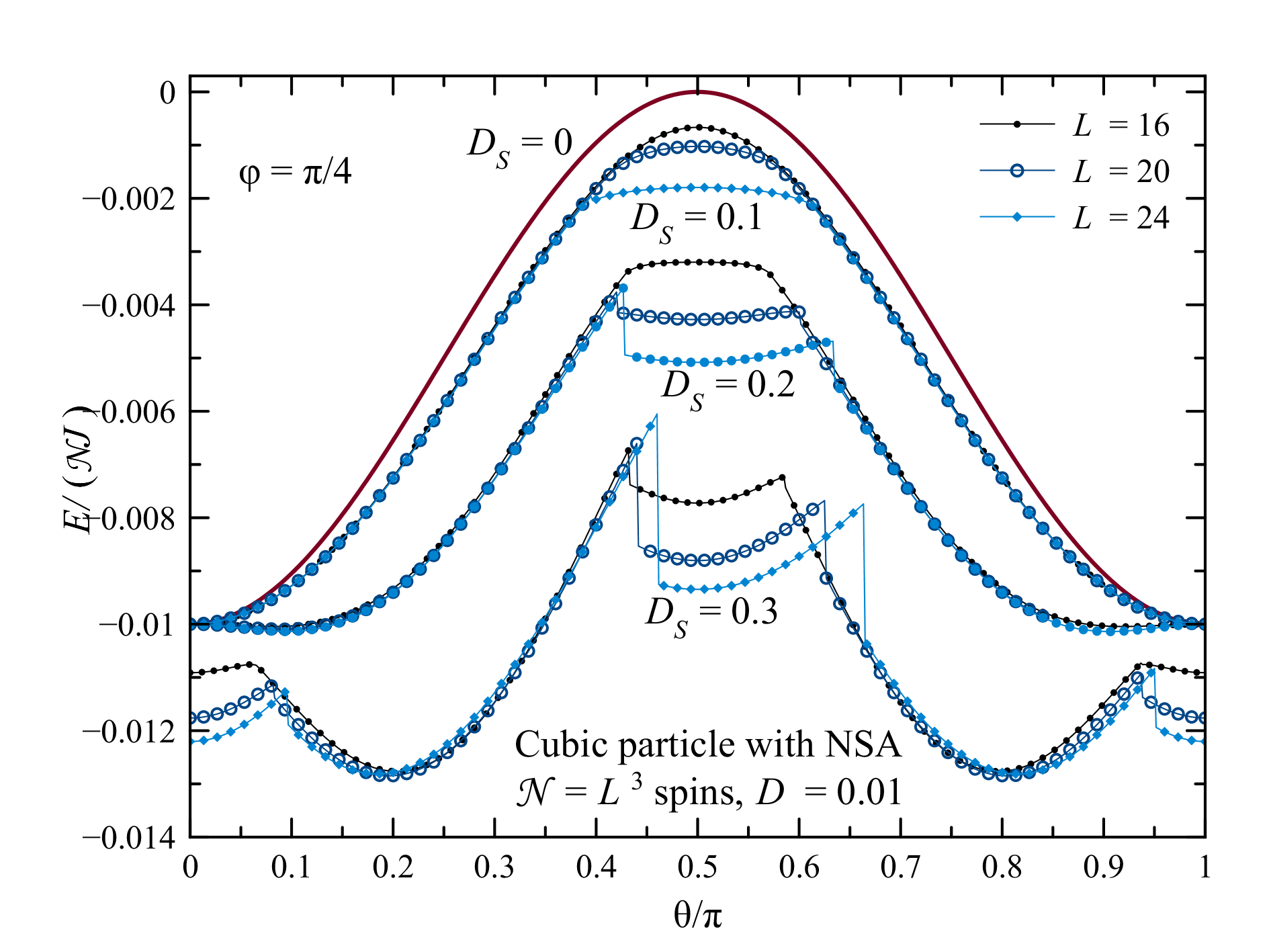}
\par\end{centering}
\caption{Energy landscapes for cubic particles with the surface and core anisotropies:
Numerical results for the particle sizes $L=16$, 20, 24 and the SA
strenghts $D_{S}=0.1,$ 0.2, 0.3. For larger $L$ and $D_{S}$, the
barrier in the middle is lowered because of the instability leading
to deviations from the single-domain state. For even larger $L$ and
$D_{S}$, the result of the constrained energy minimization becomes
multi-valued. Upper panel: $D=0$. Lower panel: $D=0.01$.}

\label{Fig-dE_vs_theta_L=00003D16_20_24}
\end{figure}

\begin{figure}
\begin{centering}
\includegraphics[width=9cm]{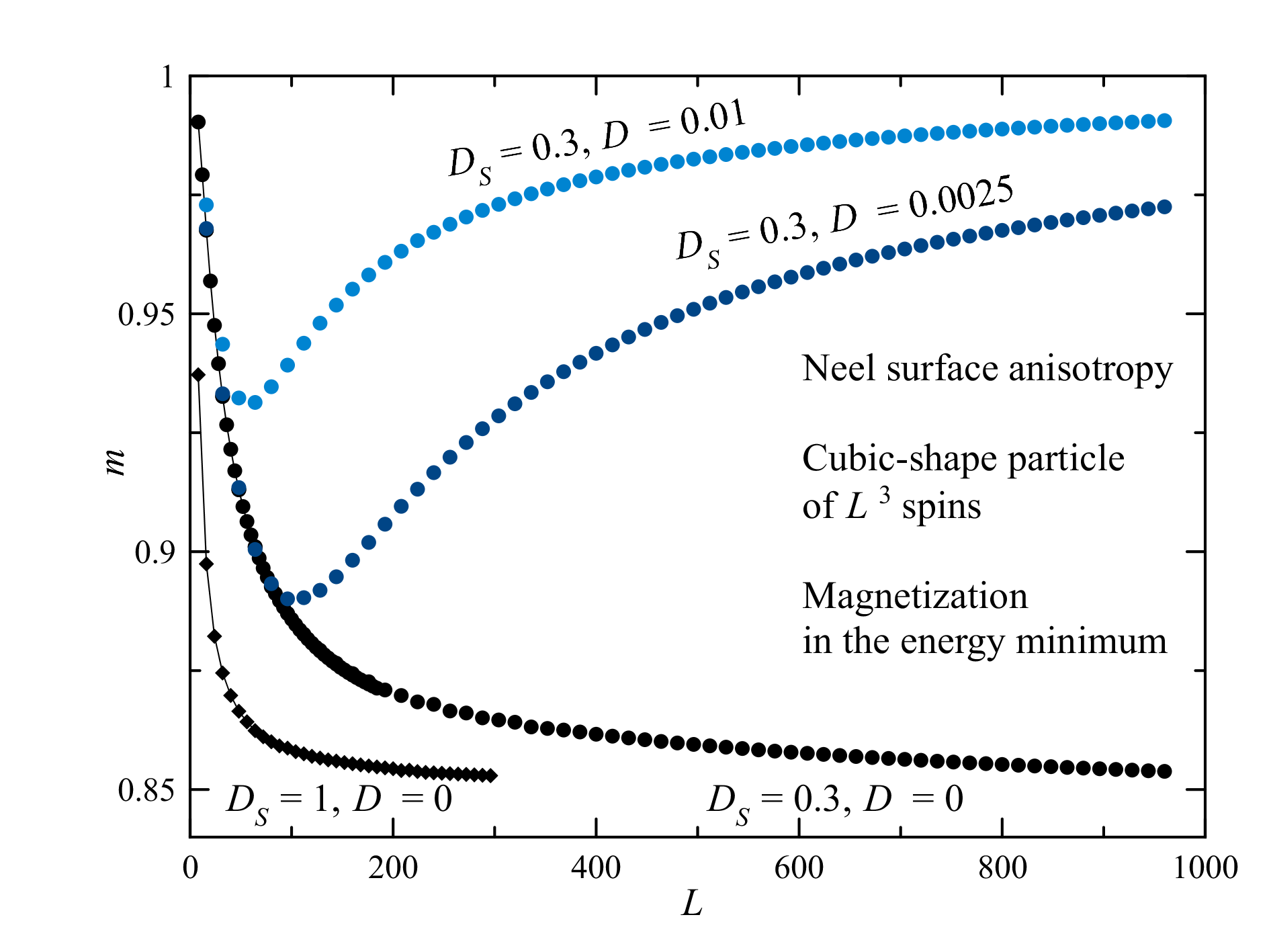}
\par\end{centering}
\begin{centering}
\includegraphics[width=9cm]{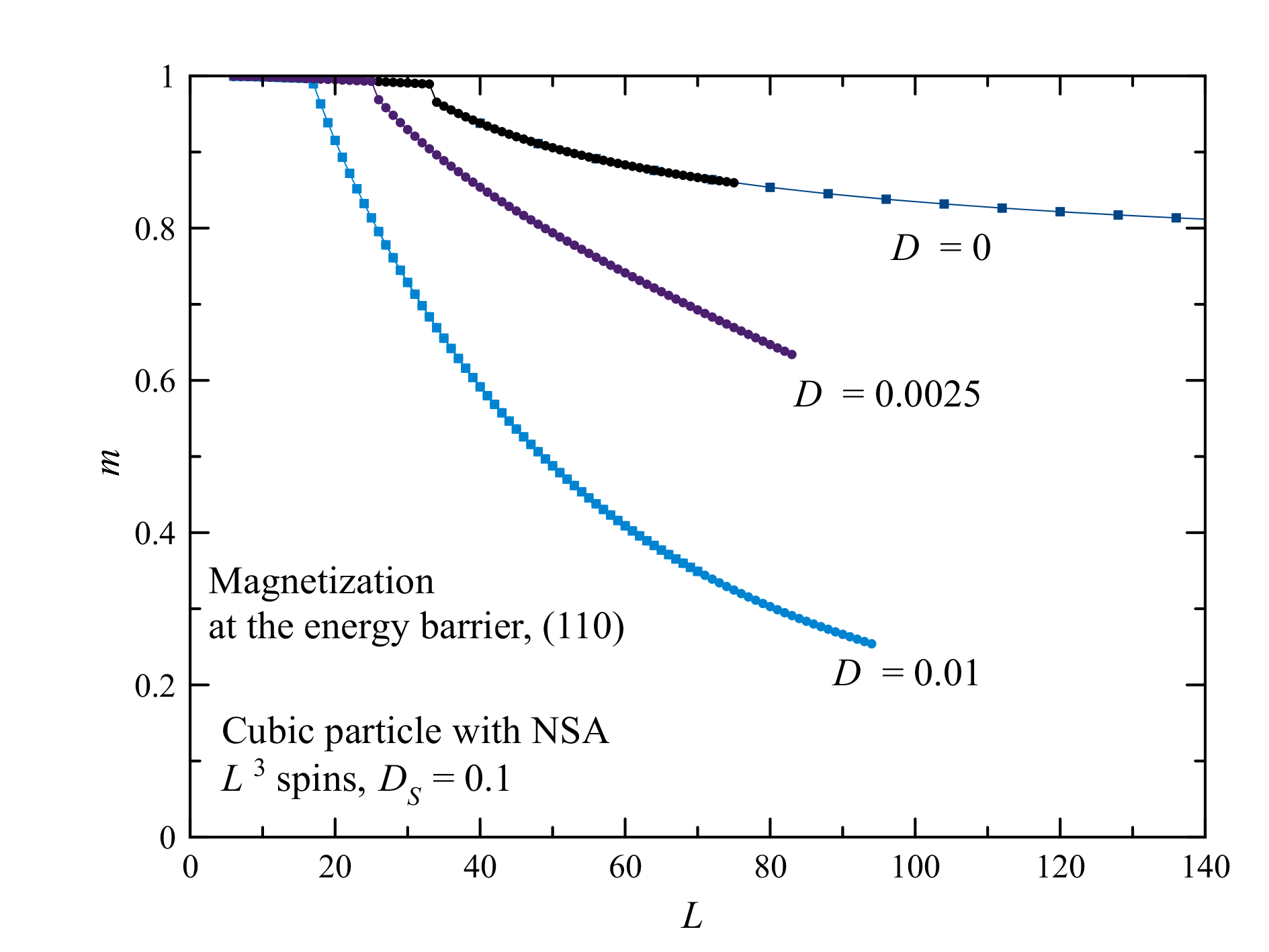}
\par\end{centering}
\caption{Particle's magnetization $m$ vs its linear size $L$ for $D_{S}=0.1$
and different values of the core anisotropy $D$. Upper panel: magnetization
in the direction of the energy minimum, $\left(1,1,1\right)$ for
$D=0$. Here $D>0$ stabilizes the quasi-collinear state by screening
the surface perturbations. Lower panel: magnetization in the direction
of the energy barrier, $\left(1,1,0\right)$. Here $D>0$ destabilizes
the quasi-collinear state because of the tendency to form a domain
wall across the particle.}

\label{Fig-m_vs_L}
\end{figure}

Fig. \ref{Fig-Ered_vs_theta_DV=00003D0} shows the energy landscapes
of cubic particles of sizes $L=16$ and 32 computed by the constrained
energy minimization as explained in Sec. \ref{sec:Numerical-methods}
together with the analytical result of Eq. (\ref{H_SA^(2)_DS_result}).
There is a fair overall agreement between the numerical and analytical
results, although Eq. (\ref{H_SA^(2)_DS_result}) shows deeper energy
minima. The discrepansy must be due to finite-size effects. Indeed,
each face contains only $(L-2)^{2}$ sites subject to the SA rather
than $L^{2}$ sites. This suggests renormalization of $D_{S}$ as
$D_{S}\Rightarrow\tilde{D}_{S}=(1-2/L)^{2}D_{S}$ that results in
the additional factor $(1-2/L)^{4}$ in $\mathcal{H}_{SA}^{(2)}$.
However, this renormalization would be too strong for the results
in Fig. \ref{Fig-Ered_vs_theta_DV=00003D0} making the energy minima
a way too shallow. However, there are edges working in the same directions
as faces, only weaker. Also the exchange interaction weakens near
the surfaces because of the missing neighbors. In the absence of an
analytical solution for the lattice problem, one can fit the finite-size
effect replacing the effective number of spins in the face by $(L-\varsigma)^{2}$.
The results of Eq. (\ref{H_SA^(2)_DS_result}) with the additional
factor $(1-\varsigma/L)^{4}$ with $\varsigma=0.7$ in $\mathcal{H}_{SA}^{(2)}$
shown in Fig. \ref{Fig-Ered_vs_theta_DV=00003D0} as ``Analytical,
finite size'' are closer to the numerical results than the pure results
of Eq. (\ref{H_SA^(2)_DS_result}) labeled ``Analytical''.

Whereas for $L=16$ the numerical results for the two different values
of $D_{S}$ scale, for $L=32$ there are visible deviations from scaling.
In particular, for $D_{S}=0.1$ the energy of the $\left(0,0,1\right)$
state is lowered due to the instability of the collinear state in
which spins near some surfaces turn by $90^{\circ}$under the influence
of the SA. This can be seen in the lower panel of Fig. \ref{Fig_spins_L=00003D32_DS=00003D0.1}.
This state cannot be obtained within the linear approximation. For
$L=16$ there is still no instability and the $\left(0,0,1\right)$
state is strictly collinear. On the other hand, the $\left(1,1,0\right)$
state in the upper panel of Fig. \ref{Fig_spins_L=00003D32_DS=00003D0.1}
is that given by Eq. (\ref{psi_solution_DS_only_cube}) and its numerically
found magnetization is $m=0.9898$ while Eq. (\ref{m_DS_result})
yields a close value $m=0.9858$. On the other hand, the magnetization
in the unstable $\left(0,0,1\right)$ state for $L=32$ is lower:
$m=0.9680$.

Numerical results for the energy landscape of prolate and oblate box-shape
particles with $D_{S}=0.1$ are shown in Fig. \ref{Fig-Ered_vs_theta_oblate_and_prolate}.
In this case there is the first-order contribution to the effective
anisotropy $\mathcal{H}_{SA}^{(1)}$ given by Eq. (\ref{H_SA_1_result_uniax}).
The second-order term can be computed as the difference: $\mathcal{H}_{SA}^{(2)}=\mathcal{H}_{\mathrm{eff}}-\mathcal{H}_{SA}^{(1)}$,
where $\mathcal{H}_{\mathrm{eff}}$ is the numerically obtained particle's
energy. One can see that the second-order term can be large enough
to compete with the first-order one. For prolate and oblate particles,
$\mathcal{H}_{SA}^{(2)}$ is very close to the cubic-particle result,
as shown in Fig. \ref{Fig-Ered_vs_theta_DS=00003D0.1_cubic_and_box}.

Fig. \ref{Fig-dE_vs_theta_L=00003D16_20_24} shows the energy landscapes
for three different particle's sizes and three different values of
$D_{S}$ for the core anisotropy $D=0$ and $D=0.01$. In contrast
to Fig. \ref{Fig-Ered_vs_theta_DV=00003D0}, the energy is shown not
scaled with $D_{S}^{2}$. One can see that for larger $L$ and $D_{S}$
the barrier in the middle is flattened and lowered because of the
instability leading to the deviation from the single-domain barrier
state with all spins perpendicular to $z$ axis. As the result of
this instability, spins on one side of the cube turn toward $z$ axis
to lower the energy, whereas spins on the other side turn in the opposite
direction \cite{gar18_uniform_eprint}. Further increasing $L$ results
in forming a domain wall in the middle of the particle, and the constrained
energy minimization fails. This state cannot be obtained within the
linear approximation. The lower panel of \ref{Fig-dE_vs_theta_L=00003D16_20_24}
shows the energy landscape dominated by the core anisotropy, however,
strongly modified by the SA. Here, too, the uniform barrier state
is destroyed for large particles and strong SA.

Dependence of the particle's magnetization on the particle's size
$L$ is shown in Fig. \ref{Fig-m_vs_L}. The role of the core anisotropy
$D$ is strikingly different for the energy-mimina and the energy-barrier
states. For $D=0$ at the minima at $\left(\pm1,\pm1,\pm1\right)$,
the magnetization deficit is growing with $L$ according to Eq. (\ref{m_DS_result}),
so $m$ goes down. However, for larger $L$ the saturation state is
reached in which the surface spins are oriented according to the SA
(perpendicular to the surfaces near the surfaces for $D_{S}>0$).
In this state, instead of Eq. (\ref{psi_solution_DS_only_cube}),
$\boldsymbol{\psi}(\mathbf{r})$ {[}or, rather, $\mathbf{s}(\mathbf{r})${]}
is a function of $\mathbf{r}/L$ only, independently of $a$. Thus
$m$ becomes a geometrical constant $m\simeq0.85$ independent of
$L$ and $D_{S}$. For $D>0$, perturbations from the surface become
screened at the distance of the domain-wall width $\delta=\sqrt{J/(2D)}$.
Thus on increasing $L$ the magnetization $m$ at first decreases
until $L\sim\delta$, then increases again because of the screening.
This is clearly seen in the upper panel of Fig. \ref{Fig-m_vs_L}
where the extremely large values of $L$ should be noticed.

In the lower panel of Fig. \ref{Fig-m_vs_L}, the magnetization in
the barrier state $\left(110\right)$ is close to 1 for $L$ small
enough, while the spin configuration is shown in the upper panel of
Fig. \ref{Fig_spins_L=00003D32_DS=00003D0.1}. Further increase of
$L$ causes instabilities of the surface spins in the $xy$ surfaces:
for $D_{S}>0$ these spins turn perpendicular to the surfaces parallel
$z$ axis. In the limit $L\rightarrow\infty$ for $D=0$, a state
with $\mathbf{s}(\mathbf{r})$ depending on $\mathbf{r}/L$ only should
be reached, in which $m$ is a another geometrical constant. However,
$D>0$ leads to the instability at smaller $L$ with the subsequent
formation of a domain wall in the middle of the particle. After that
the constrained energy minimization method fails, that's why the $D>0$
curves in the lower panel of Fig. \ref{Fig-m_vs_L} could not be computed
for larger $L$. To the contrary, $D_{S}<0$ stabilizes surface spins
in the $xy$ planes, so that the discussed instability does not happen
for $D=0$ and requires the values of $D$ exceeding some threshold
to develop.

\section{Screening and other generalizations\label{sec:Screening}}

In this section the results of Sec. \ref{sec:Cubic-particle-with}
will be generalized for the model with the uniaxial anisotropy and
magnetc field. Some calculations will be made for a parallelepiped
particle where the first-order effective surface anisotropy is present.
In the continuous approximation, the Hamiltonian has the form
\begin{equation}
\mathcal{H}=\intop\frac{dV}{a^{3}}\left[\frac{a^{2}J}{2}\left(\frac{\partial s_{\alpha}}{\partial\mathbf{r}}\right)^{2}-Ds_{z}^{2}-\mathbf{h}\cdot\mathbf{s}-\frac{D_{S}}{2}a\delta_{S}\left(\mathbf{n}\cdot\mathbf{s}\right)^{2}\right].\label{Ham_continuous_SA-1}
\end{equation}
Using Eq. (\ref{psi_Def}), from the first of equations (\ref{constrained_equations})
one obtains
\begin{equation}
0=(\boldsymbol{\nu}+\boldsymbol{\psi})\times\left[\mathbf{h}+2D\left(\nu_{z}+\psi_{z}\right)\mathbf{e}_{z}+a^{2}J\Delta\boldsymbol{\psi}+\mathbf{h}_{\lambda}\right],
\end{equation}
where $\mathbf{h}_{\lambda}$ is given by Eq. (\ref{h_constr_via_h_eff}).
The boundary conditions are defined by Eq. (\ref{psi_boundary_conditions}).
Substituting $\mathbf{h}_{\lambda}$ and rearranging keeping only
the linear-$\boldsymbol{\psi}$ terms, one arrives at
\begin{eqnarray}
\boldsymbol{\nu}\times\left[2D\psi_{z}\mathbf{e}_{z}+a^{2}J\Delta\boldsymbol{\psi}-\mathbf{h}_{SA}\right.\nonumber \\
-\left.\boldsymbol{\nu}\cdot\left(\mathbf{h}+2D\nu_{z}\mathbf{e}_{z}+\mathbf{h}_{SA}\right)\boldsymbol{\psi}\right] & = & 0.\label{psi_D_equation}
\end{eqnarray}
One can search for the solution in the form
\begin{equation}
\boldsymbol{\psi}=\psi_{1}\boldsymbol{\nu}_{1}+\psi_{2}\mathbf{\boldsymbol{\nu}}_{2},\label{psi_12_Def}
\end{equation}
where $\mathbf{\boldsymbol{\nu}}_{1}$ and $\mathbf{\boldsymbol{\nu}}_{2}$
are unit vectors perpendicular to $\boldsymbol{\nu}$ and to each
other, so that $\boldsymbol{\nu}\times\mathbf{\boldsymbol{\nu}}_{1}=\mathbf{\boldsymbol{\nu}}_{2}$.
It is convenient to choose $\mathbf{\boldsymbol{\nu}}_{1}\cdot\mathbf{e}_{z}=0$
and $\mathbf{\boldsymbol{\nu}}_{2}$ in the plane spanned by $\mathbf{e}_{z}$
and $\boldsymbol{\nu}$. Then equations for $\psi_{1}$ and $\psi_{2}$
decouple, and after some algebra one obtains Helmholtz equations with
sources
\begin{eqnarray}
\Delta\psi_{1}-\kappa_{1}^{2}\psi_{1} & = & \mathbf{\boldsymbol{\nu}}_{1}\cdot\mathbf{h}_{SA}\nonumber \\
\Delta\psi_{2}-\kappa_{2}^{2}\psi_{2} & = & \mathbf{\boldsymbol{\nu}}_{2}\cdot\mathbf{h}_{SA},\label{psi_12_eqs}
\end{eqnarray}
where
\begin{eqnarray}
\kappa_{1}^{2} & \equiv & \frac{\mathbf{\boldsymbol{\nu}}\cdot\left(\mathbf{h}+\mathbf{h}_{SA}\right)+2D\nu_{z}^{2}}{a^{2}J}\nonumber \\
\kappa_{2}^{2} & \equiv & \frac{\mathbf{\boldsymbol{\nu}}\cdot\left(\mathbf{h}+\mathbf{h}_{SA}\right)+2D\left(2\nu_{z}^{2}-1\right)}{a^{2}J}.\label{k_12_result}
\end{eqnarray}
Here $\kappa^{2}>0$ corresponds to the exponentially decaying perturbations
(screening), whereas $\kappa^{2}<0$ describes proliferating perturbations
(anti-screening). For instance, for $h=h_{SA}=0$ and $\nu_{z}>1/\sqrt{2}$
$\left(\theta<\pi/4\right)$ both $\kappa_{1}^{2}$ and $\kappa_{2}^{2}$
are positive and the uniaxial anisotropy stabilizes the particle's
magnetization. Larger deviations from the easy axis lead to $\kappa_{2}^{2}<0$
and destruction of the particle's magnetization. For the source terms
in the case $N_{x}=N_{y}\equiv N_{\bot}$ from Eq. (\ref{h_SA_uniaxial})
one obtains
\begin{equation}
\mathbf{\boldsymbol{\nu}}_{1}\cdot\mathbf{h}_{SA}=0,\qquad\mathbf{\boldsymbol{\nu}}_{2}\cdot\mathbf{h}_{SA}=-2D_{S}\frac{N_{\bot}-N_{z}}{N_{\bot}N_{z}}\nu_{z}\sqrt{1-\nu_{z}^{2}}.
\end{equation}

In the sequel, we consider cubic particles for which $\mathbf{h}_{SA}=0$.
The solution of Eqs. (\ref{psi_12_eqs}) can be searched for in the
form
\begin{equation}
\psi_{\alpha}=\frac{D_{S}}{aJ}\frac{C_{\alpha x}\cosh\kappa_{\alpha}x+C_{\alpha y}\cosh\kappa_{\alpha}y+C_{\alpha z}\cosh\kappa_{\alpha}z}{\kappa_{\alpha}\sinh\left(\kappa_{\alpha}L/2\right)},\label{psi_alpha_result_kappa}
\end{equation}
where $\alpha=1,2$ and $\boldsymbol{\psi}$ is defined by Eq. (\ref{psi_12_Def}).
This function satisfies the Helmholtz equations, if the sum of the
$C$-coefficients is zero. They can be determined from the boundary
conditions, Eq. (\ref{psi_boundary_conditions}). Using Eq. (\ref{f_on_faces}),
one obtains
\begin{eqnarray}
C_{\alpha x} & = & \left.f_{\alpha}\left(\boldsymbol{\nu},\mathbf{n}\right)\right|_{x=L/2}=\nu_{x}\nu_{\alpha x}\nonumber \\
C_{\alpha y} & = & \left.f_{\alpha}\left(\boldsymbol{\nu},\mathbf{n}\right)\right|_{y=L/2}=\nu_{y}\nu_{\alpha y}\nonumber \\
C_{\alpha z} & = & \left.f_{\alpha}\left(\boldsymbol{\nu},\mathbf{n}\right)\right|_{z=L/2}=\nu_{z}\nu_{\alpha z},\label{C_alpha_xyz_result}
\end{eqnarray}
where $\nu_{\alpha x}=\boldsymbol{\nu}_{\alpha}\cdot\mathbf{e}_{x}$,
etc. One can see that
\begin{equation}
C_{\alpha x}+C_{\alpha y}+C_{\alpha z}=\boldsymbol{\nu}\cdot\boldsymbol{\nu}_{\alpha}=0,
\end{equation}
as it should be. With the current choice of the vectors $\mathbf{\boldsymbol{\nu}}_{1}$
and $\mathbf{\boldsymbol{\nu}}_{2}$, the explicit form of the $C$-coefficients
is
\begin{equation}
C_{1x}=-C_{1y}=\frac{\nu_{x}\nu_{y}}{\sqrt{1-\nu_{z}^{2}}},\qquad C_{1z}=0,\label{C_1_xyz_result}
\end{equation}
and
\begin{equation}
C_{2x}=C_{2y}=\frac{\nu_{y}^{2}\nu_{z}}{\sqrt{1-\nu_{z}^{2}}},\qquad C_{2z}=-\nu_{z}\sqrt{1-\nu_{z}^{2}}.\label{C_2_xyz_result}
\end{equation}

In the case $\kappa=ik$ (the anti-screening case), Eq. (\ref{psi_alpha_result_kappa})
becomes
\begin{equation}
\psi_{\alpha}=-\frac{D_{S}}{aJ}\frac{C_{\alpha x}\cos k_{\alpha}x+C_{\alpha y}\cos k_{\alpha}y+C_{\alpha z}\cos k_{\alpha}z}{k_{\alpha}\sin\left(k_{\alpha}L/2\right)}.\label{psi_alpha_result-k}
\end{equation}
At $k_{\alpha}\rightarrow0$ this expression is regular but it diverges
at $k_{\alpha}L\rightarrow2\pi$. For instance, for $h=0$ and $\nu_{z}=0$
in Eq. (\ref{k_12_result}) one has $k_{2}=1/\delta,$ so that the
particle's size should satisfy $L<2\pi\delta$. However, in the model
with a uniaxial ansotropy there is another stability criterion \cite{gar18_uniform_eprint},
$L<\pi\delta$, for the same state with the spin perpendicular to
the easy axis \textendash{} the barrier state. If this condition is
violated, then there is a finite $\psi$ even in the absence of the
surface anisotropy. Thus, the divergence of the solution at $L=2\pi\delta$
is beyond the applicability range of the linearization method. For
$D=0$, there is no corresponding instability, but screening and antiscreening
can be created by the magnetic field. In this case, the point $k_{\alpha}L=2\pi$
can be approached, and this defines the applicability of the method.

The energy of the particle at second order in $\boldsymbol{\psi}$
is given by Eq. (\ref{E_psi_continuous}) with the additional term
$-D\psi_{z}^{2}$ in square brackets. The terms linear in $\boldsymbol{\psi}$
vanish because of Eq. (\ref{psi_sum_rule}). After some algebra one
arrives at the final result
\begin{eqnarray}
\mathcal{H}_{\mathrm{eff}} & = & -\mathcal{N}\left(D\nu_{z}^{2}+\mathbf{h}\cdot\boldsymbol{\nu}\right)-\frac{\mathcal{N}D_{S}^{2}}{3J}\left[\frac{\nu_{x}^{2}\nu_{y}^{2}}{1-\nu_{z}^{2}}F\left(\kappa_{1}L\right)\right.\nonumber \\
 &  & +\left.\left(\nu_{z}^{2}\left(\nu_{x}^{2}+\nu_{y}^{2}\right)-\frac{\nu_{x}^{2}\nu_{y}^{2}\nu_{z}^{2}}{1-\nu_{z}^{2}}\right)F\left(\kappa_{2}L\right)\right]\nonumber \\
 &  & -\frac{L^{2}}{\delta^{2}}\frac{\mathcal{N}D_{S}^{2}}{J}\nu_{z}^{2}\left(\nu_{x}^{4}+\nu_{y}^{4}+\nu_{x}^{2}\nu_{y}^{2}\right)F_{D}\left(\kappa_{2}L\right),\label{E_full_result-2}
\end{eqnarray}
where $\delta=a\sqrt{J/(2D)}$,
\begin{equation}
F(x)=\frac{3}{x}\frac{3\sinh\left(x\right)+x}{\cosh\left(x\right)-1}-\frac{24}{x^{2}}\cong\begin{cases}
1-\frac{x^{4}}{2520}, & x\ll1\\
\frac{9}{x}-\frac{24}{x^{2}}, & x\gg1
\end{cases}
\end{equation}
and
\begin{equation}
F_{D}(x)=\frac{1}{x^{3}}\frac{\sinh\left(x\right)+x}{\cosh\left(x\right)-1}-\frac{4}{x^{4}}\cong\begin{cases}
\frac{1}{180}-\frac{x^{2}}{3780}, & x\ll1\\
\frac{1}{x^{3}}-\frac{4}{x^{4}}, & x\gg1.
\end{cases}.
\end{equation}
In the case of $\kappa=ik$ one has to replace $F\left(ikL\right)\Rightarrow G(kL),$
where
\begin{equation}
G(x)=\frac{24}{x^{2}}-\frac{3}{x}\frac{3\sin\left(x\right)+x}{1-\cos\left(x\right)}
\end{equation}
has the same behavior as $F(x)$ at $x\ll1$ but diverges at $x=2\pi$.
The last term in Eq. (\ref{E_full_result-2}) is the cross-term originating
from $-D\psi_{z}^{2}$ in the integrand of the energy.

For $D=0$, one has $\kappa_{1}=\kappa_{2}=\kappa$, and the energy
simplifies to
\begin{eqnarray}
\mathcal{H}_{\mathrm{eff}} & = & -\mathcal{N}\mathbf{h}\cdot\boldsymbol{\nu}-\frac{\mathcal{N}D_{S}^{2}}{3J}F\left(\frac{L}{a}\sqrt{\frac{\mathbf{h}\cdot\boldsymbol{\nu}}{J}}\right)\nonumber \\
 &  & \qquad\times\left(\nu_{x}^{2}\nu_{y}^{2}+\nu_{y}^{2}\nu_{z}^{2}+\nu_{z}^{2}\nu_{x}^{2}\right).
\end{eqnarray}
Here there can be screening or anti-screening because of the magnetic
field. For $h=0$, Eq. (\ref{H_SA^(2)_DS_result}) is recovered.

In the limit of $\kappa_{1}L,\kappa_{2}L\ll1$, the second-order part
of Eq. (\ref{E_full_result-2}) simplifies to Eq. (\ref{H_SA^(2)_DS_result}).
In this case, the first- and second-order terms in the effective particle's
anisotropy are additive. The leading correction term of order $\left(L/\delta\right)^{2}$
comes from the cross-term in the energy with a small numerical factor.
The corrections from the terms with the function $F$ are of order
$\left(L/\delta\right)^{4}$ with an extremely small numerical factor.

\begin{figure}
\begin{centering}
\includegraphics[width=9cm]{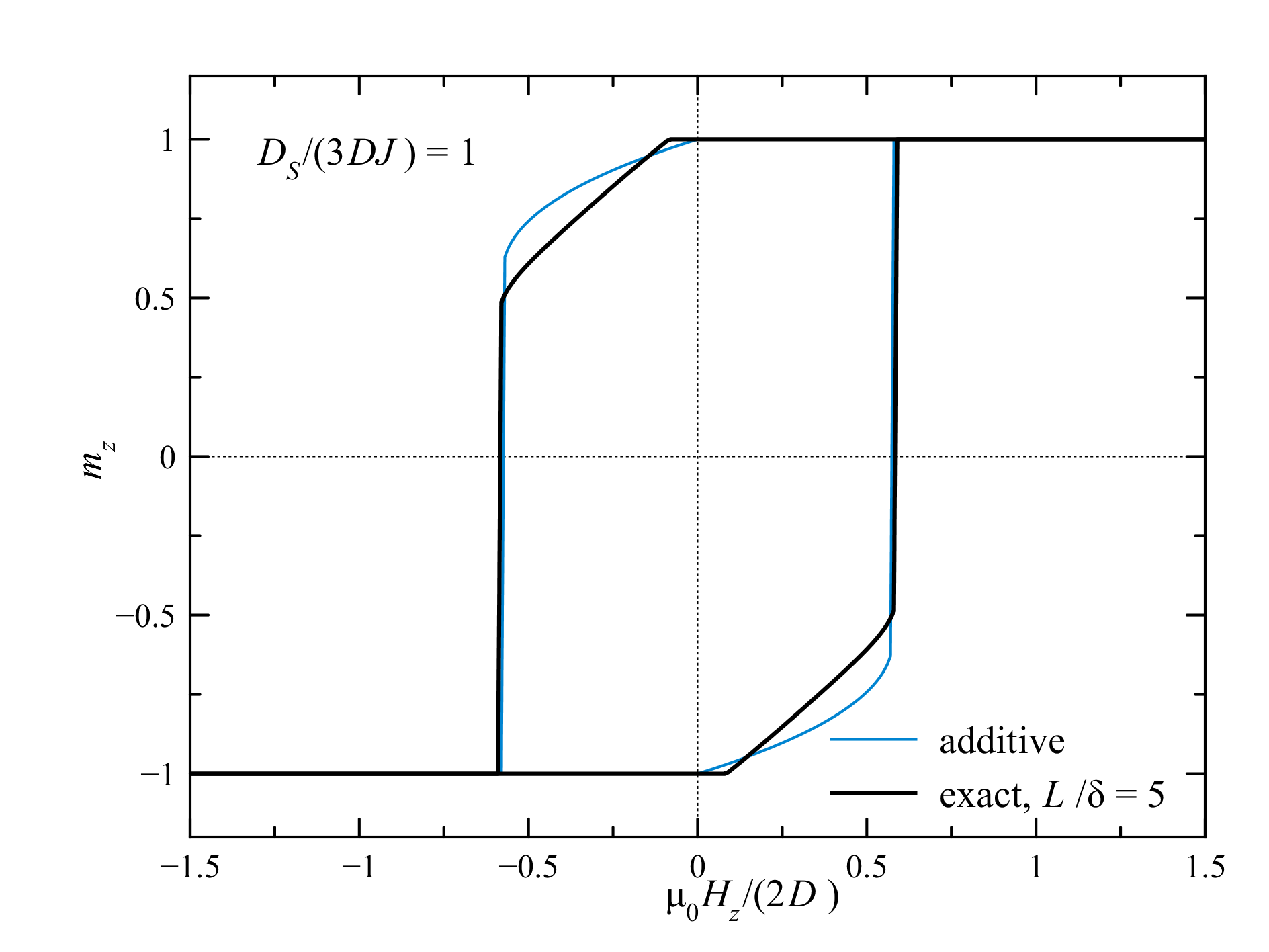}
\par\end{centering}
\begin{centering}
\includegraphics[width=9cm]{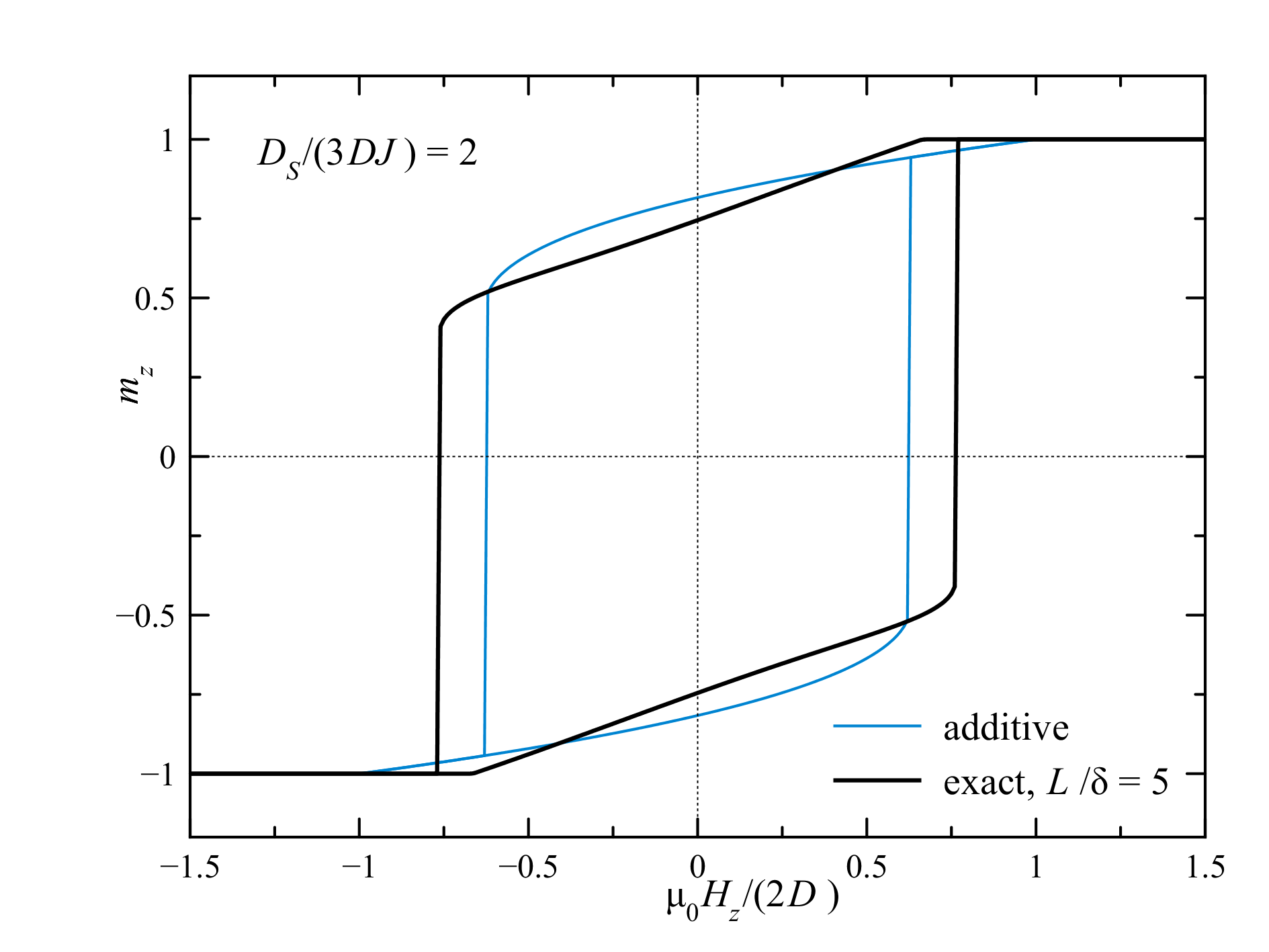}
\par\end{centering}
\caption{Hysteresis loops for $\mathbf{H}$ along $z$ axis, using the additive
effective particle's Hamiltonian, Eq. (\ref{Ham_eff_1+2}), and the
exact analytical solution, Eq. (\ref{E_full_result-2}), that depends
on the ratio $L/\delta$.}

\label{Fig_mz_vs_h}
\end{figure}

Energy landscapes plotted using Eq. (\ref{E_full_result-2}) for $L$
small enough show very small deviations from the results obtained
using the additive effective Hamiltonian, Eq. (\ref{Ham_eff_1+2}).
For larger $L$, some deviations are seen but then, with the further
inclease of $L$, the solution quickly diverges for the orientations
having the imaginary $\kappa$ \textendash{} near the barriers and
opposite to the magnetic field, where anti-screening occurs. As an
example, hysteresis loops for $\mathbf{H}$ along $z$ axis are shown
in Fig. \ref{Fig_mz_vs_h}. The results of the additive model do not
depend on $L/\delta$. The exact results using Eq. (\ref{E_full_result-2})
are very close to the latter for $L/\delta\lesssim5$. However, for
$L/\delta\gtrsim5$ the solution diverges and the hysteresis loop
breaks down.

For the field along $z$ axis, the energy of the particle can be minimized
with respect to the azimuthal angle $\varphi$ that yields $\varphi=\pi/4$
and equivalent solutions. For these values of $\varphi$, one can
write down the compact expression of the energy in terms of $x\equiv\cos\theta$.
Eq. (\ref{E_full_result-2}) in the reduced form becomes
\begin{eqnarray}
e(x) & = & -x^{2}-2\alpha x-\frac{1}{4}\beta\left(1-x^{2}\right)\left[F\left(\kappa_{1}L\right)+3x^{2}F\left(\kappa_{2}L\right)\right]\nonumber \\
 &  & \qquad-\frac{9}{4}\beta x^{2}\left(1-x^{2}\right)^{2}\tilde{L}^{2}F_{D}\left(\kappa_{2}L\right),
\end{eqnarray}
where
\begin{equation}
e\equiv\frac{\mathcal{H}_{\mathrm{eff}}}{\mathcal{N}D},\quad\alpha\equiv\frac{h}{2D},\quad\beta\equiv\frac{D_{S}^{2}}{3DJ},\quad\tilde{L}\equiv\frac{L}{\delta}
\end{equation}
and
\begin{equation}
\kappa_{1}L=\tilde{L}\sqrt{\alpha x+x^{2}},\qquad\kappa_{2}L=\tilde{L}\sqrt{\alpha x+2x^{2}-1}.
\end{equation}

In the case of zero field and dominating uniaxial anisotropy, the
dimensionless energy barrier is given by
\begin{equation}
u=e(0)-e(1)=1-\frac{1}{4}\beta.
\end{equation}
It does not depend on screening and is the same as within the additive
approximation. To investigate the stability of the state along $z$
axis, $x=1$, one can expand $e(x)$ in terms of $\delta x\equiv1-x$.
This yields
\begin{equation}
e\cong-1-2\alpha+2\left[1+\alpha-\beta F\left(\tilde{L}\sqrt{1+\alpha}\right)\right]\delta x
\end{equation}
thus the energy minimum is stable for
\begin{equation}
\frac{1+\alpha}{F\left(\tilde{L}\sqrt{1+\alpha}\right)}>\beta.
\end{equation}
For small particles screening is negligible, $F\cong1$, and one obtains
the condition $1+\alpha>\beta$. For large particles, one uses the
asymptotic form $F(x)\cong9/x$ that results in the condition $\left(1+\alpha\right)^{3/2}>9\beta/\tilde{L}$
that means a greater stability against the surface effects parametrized
by $\beta$. In the upper panel of Fig. \ref{Fig_mz_vs_h}, $\beta=1$,
so that within the additive approximation the energy minimum $x=1$,
i.e., $m_{z}=1$ exists for $\alpha>0$, i.e., $H>0$. Screening in
the exact solution makes this energy minimum more stable, so that
it disappears at the negative field corresponding to $\alpha=\left(9\beta/\tilde{L}\right)^{2/3}-1$,
as can be seen in Fig. \ref{Fig_mz_vs_h}. These results are also
related to precession frequencies near energy minima and can be important
for the magnetic resonance in magnetic nanoparticles \cite{kacsch07epjb,basvergarkac17jpcm}.

\section{Thermally-activated magnetization switching\label{sec:Thermally-activated-escape}}

\begin{figure}
\begin{centering}
\includegraphics[width=9cm]{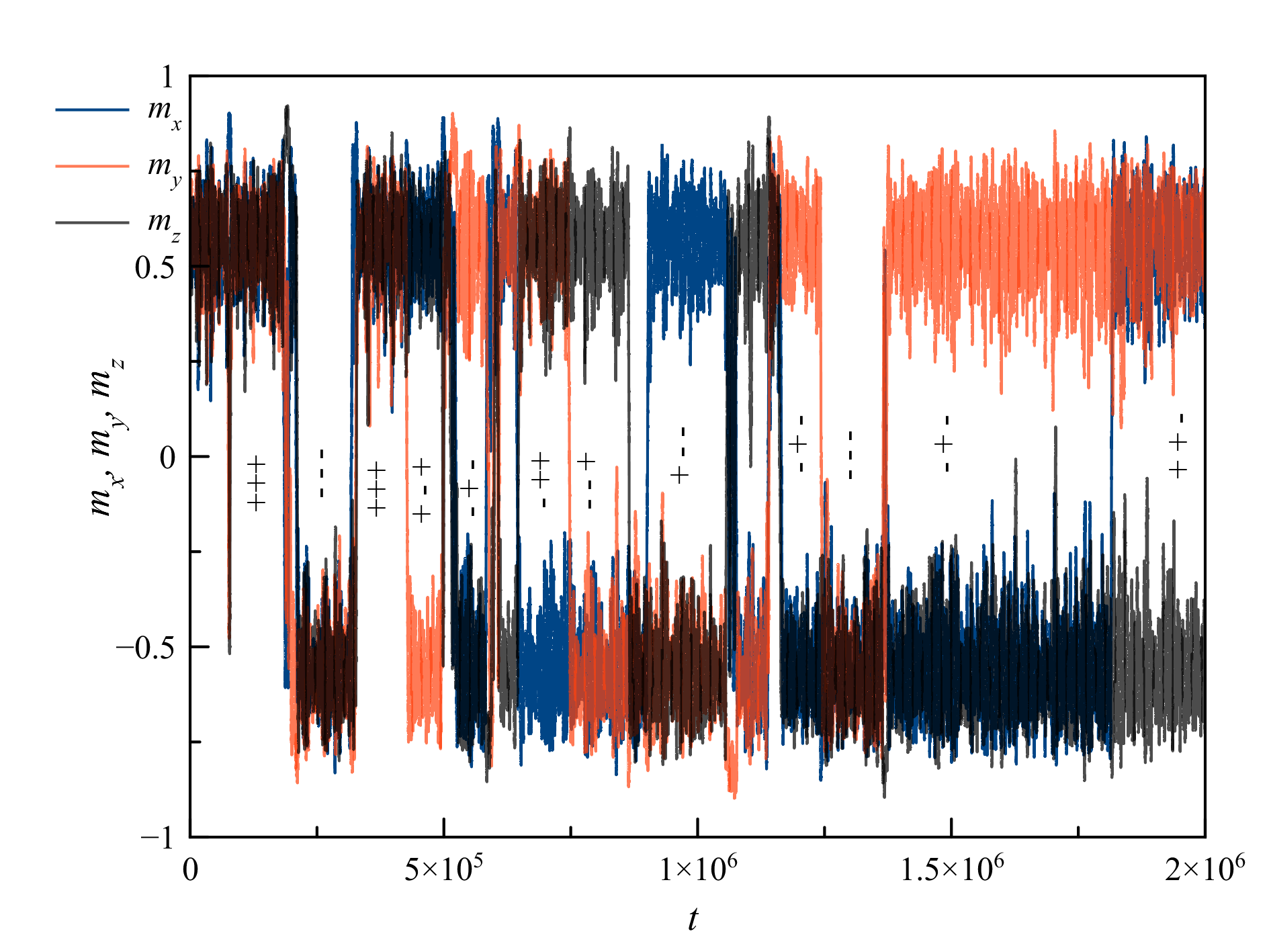}
\par\end{centering}
\begin{centering}
\includegraphics[width=9cm]{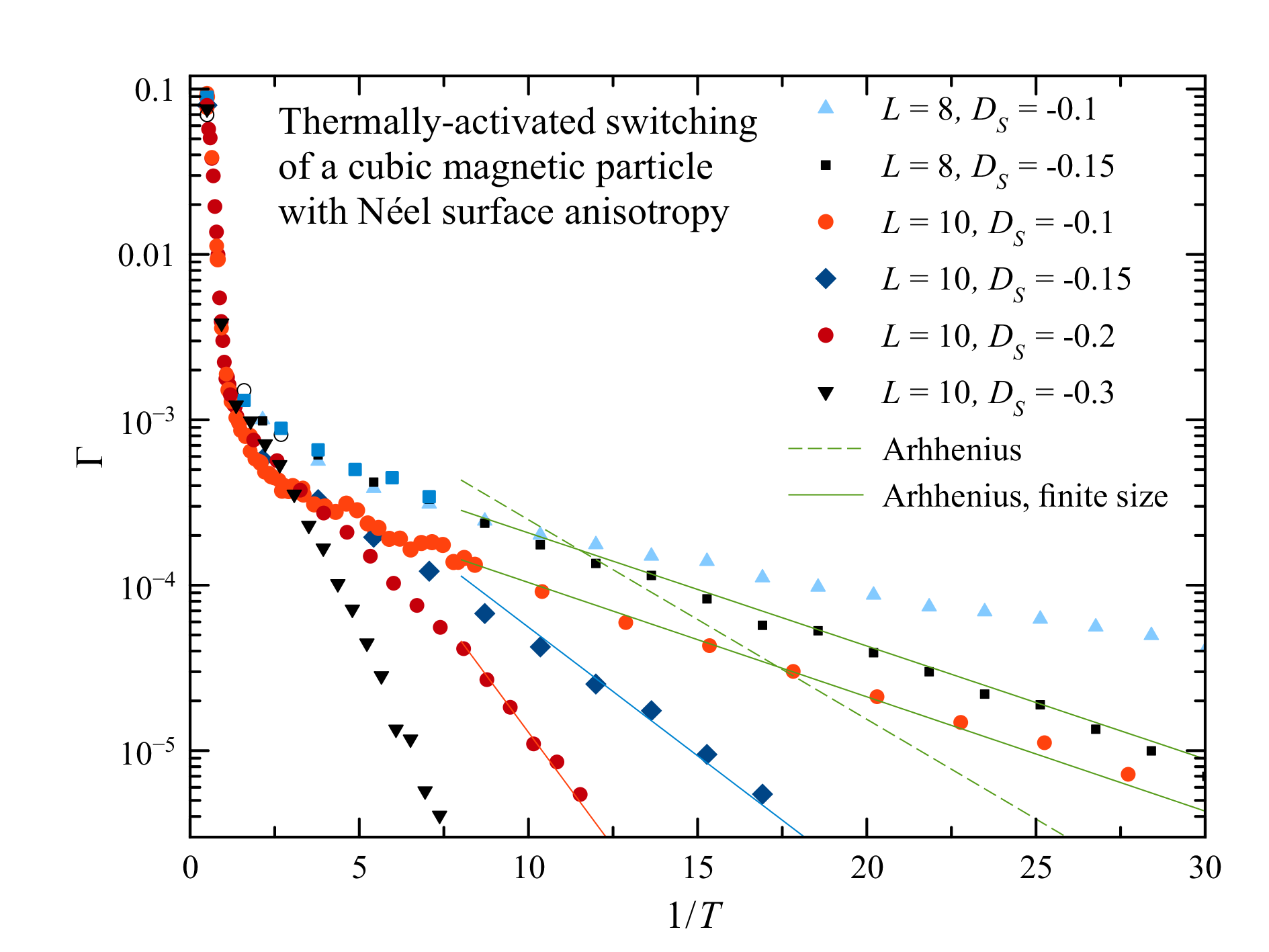}
\par\end{centering}
\caption{Thermally-activated switching of a magnetic particle with SA, considered
as a many-spin system. Upper panel: An example of the time dependence
of the particle's magnetization components for a $4\times4\times4$
cube with $D_{S}=-0.1$ at $T=0.002$. Lower panel: Switching rate
$\Gamma$ vs. the temperature for different particle's sizes and $D_{S}$
values. All computations were done with the Landau-Lifshitz damping
constant $\lambda=0.1$.}

\label{Fig_Thermal}
\end{figure}

At low temperatures, the particle spends much time in the vicinity
of the energy minima, making seldom switching to other energy minima
over energy barriers. The characteristic time of the magnetization
switching is important, for instance, for memory storage applications.
The theory gives the Arrhenuis thermal activation law for the escape
rate,
\begin{equation}
\Gamma=\Gamma_{0}e^{-U/T},
\end{equation}
where $U$ is the energy barrier. In the case of a cubic particle
with the surface anisotropy only in zero field, the barrier between
the energy minima at $\left(1,1,1\right)$ and $\left(1,1,-1\right)$
is at $\left(1,1,0\right)$, and the value of the energy barrier following
from Eq. (\ref{H_SA^(2)_DS_result}) is given by
\begin{equation}
U=\frac{\mathcal{N}D_{S}^{2}}{36J}.\label{Barrier}
\end{equation}

Switching rates for the additive core-surface effective anisotropy
of magnetic particles within the single-spin model were calculated
analytically in Ref. \cite{dejkackal08jpd} and analytically and numerically
in Ref. \cite{cofdejkal09prb}. The ac susceptibility of assemblies
of magnetic particles taking into account the effective cubic surface
anisotropy and dipolar interaction between the particles was studied
in Ref. \cite{versabkac14prb}.

To test the predictions above for the simplest case of a particle
with the SA only, considered as a many-particle system, computations
using the recently proposed pulse-noise method \cite{gar17pre} of
solving the stochastic Landau-Lifshitz equation for a system of classical
spins have been performed on cubic particles of cubic shape. This
method replaces a quasi-continuous random field by equidistant pulses
rotating all spins by random angles around random axes. Between the
pulses, the deterministic Landau-Lifshitz equation is solved with
an efficient high-order differential-equation solver. The overall
speed of this method is defined by the latter, so the method is fast
and suitable for computing on many-spin systems. Although the values
of $D_{S}$ in these computations are negative, it does not matter
because the effect of $D_{S}$ is quadratic. Switching was detected
when any of the three magnetization components changed its sign.

The results are shown in Fig. \ref{Fig_Thermal}. In the upper panel,
jumping of the magnetization of a $4\times4\times4$ cube between
the eight energy minima at a very low temperature is shown via the
three magnetization components. The behavior is typical for the strong
Lanadau-Lifshitz damping $\lambda=0.1$ used in these computations.
The escape rates were computed with the method explained in the appendix
to Ref. \cite{gar18_uniform_eprint} for the particle's sizes $8\times8\times8$
and $10\times10\times10$ and different values of $D_{S}$. The results
shown in the lower panel of Fig. \ref{Fig_Thermal} are in a fair
accord with the theory, although the the barriers given by Eq. (\ref{Barrier})
and shown by the dotted line for the $10\times10\times10$ particle
with $D_{S}=-0.1$ are too high. In fact, because of finite-size effects
the barrier given by Eq. (\ref{Barrier}) should be lower, as discussed
in Sec. \ref{sec:Numerical-results}. Here, replacing $D_{S}\Rightarrow\tilde{D}_{S}=(1-1.3/L)^{2}D_{S}$
corrects the barriers, as shown by the solid Ahhhenius lines with
fitted prefactors $\Gamma_{0}$, as even small temperature dependence
of the barrier strongly affects the prefactor and makes comparison
with the theory for the latter hardly possible. Even without these
corrections, one can see that the theory works comparing the slopes
of the temperature dependence for $L=10$, $D_{S}=-0.1$ and $L=8$,
$D_{S}=-0.15$. As the product $L^{3}D_{S}^{2}$ is nearly the same
in both cases, the barriers should be nearly the same, that is indeed
so, as can be seen in the figure.

\section{Discussion}

The cubic magnetic particle turned to be an easier object than the
spherical particle for analytically calculating the second-order effective
surface anisotropy since the linearized Laplace and Helmholtz equations
for the deviations from the collinearity can be solved directly without
using Green's functions. This is, probably, a matter of luck since
the analytical solution found for the cube cannot be easily generalized
for a parallelepiped. On the other hand, the solution for the parallelepiped
should be close to that for the cube as the numerically computed effective
particle's energy is practically independent of the particle's aspect
ratio, see Fig. \ref{Fig-Ered_vs_theta_DS=00003D0.1_cubic_and_box}.

The analytical solution found here allows to study the effect of screening
of the surface perturbations at the distances of the domain-wall width
$\delta$ in the presence of the uniaxial core anisotropy in the whole
range of $L/\delta$, where $L$ is the particle's linear size. These
results are useful near the energy minima, where screening increases
their stability. On the other hand, closer to the energy barriers
screening is replaced by the anti-screening that leads to the instability
of the linearized solution found here. It was shown that for small
and moderate $L/\delta$ the effect of screening is very small, so
that the applicability range of the additive approximation for the
terms in the effective anisotropy is rather broad.

Magnetic particle of a cubic shape can be an analytically solvable
model for other types of crystal lattices. It would be worth to investigate
whether the sign of the effective cubic anisotropy is opposite for
the fcc lattice, as has been found numerically for the spherical particles
\cite{yanetal07prb}.

Another possible extension is analytically solving the discrete problem
on the lattice instead of the Laplace equation in the continuous approximation
since for small particles the finite-size effects are quite pronounced.
\begin{acknowledgments}
This work has been supported by Grant No. DE-FG02-93ER45487 funded
by the US Department of Energy, Office of Science.
\end{acknowledgments}

\bibliographystyle{apsrev4-1}
\bibliography{../../../../../BIBLIOTHEK/gar-own,../../../../../BIBLIOTHEK/gar-books,../../../../../BIBLIOTHEK/gar-relaxation,../../../../../BIBLIOTHEK/gar-spin,../../../../../BIBLIOTHEK/gar-superparamagnetic,../../../../../BIBLIOTHEK/gar-oldworks,../../../../../BIBLIOTHEK/gar-general,C:/BIBLIOTHEK/gar-surface-nano}

\end{document}